\documentclass[11pt]{article}
\usepackage{customcommands}
\usepackage{xr-hyper}

\title{Simple Inference on Functionals of Set-Identified Parameters Defined by Linear Moments\footnotetext{A previous version of this paper was circulated under the title ``Inference on Functionals of Set-Identified Parameters Defined by Convex Moments." We are grateful to Ivan Canay, the associate editor, and two referees for excellent feedback that greatly improved the paper. We thank Victor Aguirregabiria, Bulat Gafarov, Christian Gourieroux, Jiaying Gu, Ismael Mourifie, Jeffrey Negrea, Adam Rosen, Brennan Thompson, Stanislav Volgushev and Yuanyuan Wan for helpful comments and discussion. We are also grateful to participants at the 7th Annual Doctoral Workshop in Applied Econometrics at the University of Toronto, as well as participants at the 2019 North America Summer Meeting of the Econometric Society at the University of Washington. This research was supported by the Social Sciences and Humanities Research Council of Canada. All errors are our own.}\\~\\
\vspace{-.4in}
}
\date{\small \today}
\author{JoonHwan Cho\footnote{JoonHwan Cho, Department of Economics, University of Toronto, 150 St. George Street, Toronto, Ontario, M5S3G7, Canada. Email: joonhwan.cho@mail.utoronto.ca} \\ \textit{University of Toronto}
\and 
Thomas M. Russell\footnote{Thomas M. Russell, Department of Economics, Carleton University, 1125 Colonel By Drive, Ottawa, Ontario, K1S5B6, Canada. Email: thomas.russell3@carleton.ca. 
}\\\textit{Carleton University}}

\begin{document}
\maketitle
\vspace{-.4in}
\begin{abstract}
\noindent This paper proposes a new approach to obtain uniformly valid inference for linear functionals or scalar subvectors of a partially identified parameter defined by linear moment inequalities. The procedure amounts to bootstrapping the value functions of randomly perturbed linear programming problems, and does not require the researcher to grid over the parameter space. The low-level conditions for uniform validity rely on genericity results for linear programs. The unconventional perturbation approach produces a confidence set with a coverage probability of $1$ over the identified set, but obtains exact coverage on an outer set, is valid under weak assumptions, and is computationally simple to implement.  
\end{abstract}

\medskip


\thispagestyle{empty}

\clearpage

\section{Introduction}

There is now a sizeable literature on inference procedures for partially identified models. However, many existing procedures require the researcher to perform a hypothesis test at every point in the parameter space in order to construct a confidence set, and so quickly become computationally prohibitive. Simple inference procedures familiar to applied researchers from the literature on point-identified models are also not easily transportable to the partial-identification setting due to the well-known uniformity issues that arise in these models.    

This paper proposes a different approach to obtain uniformly valid inference for linear functionals or scalar subvectors of a partially identified parameter vector in models with linear moment inequalities. Instead of explicitly inverting a hypothesis test to construct a confidence set, the proposed procedure solves a certain set of bootstrap linear programs. We show that by perturbing the problem with a small amount of random noise, a simple nonparametric bootstrap procedure can become a uniformly valid method of confidence set construction. The unconventional approach to the problem has a number of interesting consequences. The benefits are that it is fast, conceptually simple, and valid under weak assumptions. The drawback is that the resulting confidence set has a coverage probability tending to 1 over the identified set, although it has exact coverage on an outer set under some additional assumptions. In this sense, the approach modifies the coverage objective in exchange for some practical benefits. 

Our main theoretical framework builds on results from the Operations Research literature (e.g. \cite{shapiro1990concepts}, \cite{shapiro1991asymptotic}) to approximate the distribution of the value functions of linear programs with data-dependent constraints using a functional delta method. Bounding a linear functional over an identified set defined by linear moment inequalities amounts to solving two linear optimization problems: one maximization problem for the upper bound, and one minimization problem for the lower bound. The endpoints of the identified set for a linear functional of a partially identified parameter can thus be viewed as value functions of two linear programs. Under some conditions, the value functions of linear programs are known to be Hadamard differentiable with respect to the underlying data generating process. However, this form of differentiability only obtains under restrictive assumptions, and is not sufficient to construct \textit{uniformly} valid confidence sets for a partially identified linear functional or scalar subvector.\footnote{This result is reminiscent of \cite{kasy2019uniformity}, who emphasizes that failures of uniformity often result as failures of uniform versions of the delta method.}

In this paper we demonstrate the conditions under which the value functions of a linear program are \textit{uniformly Hadamard differentiable} with respect to the underlying data generating process. The conditions required for this level of differentiability include nonempty interior of the feasible region/identified set, and the existence and uniqueness of both optimal solutions and Lagrange multipliers in the underlying linear programs. 

Unfortunately, the conditions required for uniform Hadamard differentiability are high-level and difficult to verify. However, \cite{spingarn1979generic} use \textit{Sard's Theorem} from differential topology to demonstrate that a generic linear program has both a unique solution and unique Lagrange multipliers. In particular, by introducing small random perturbations to the constraints and objective function of a linear program we can ensure a unique solution and unique Lagrange multipliers with probability 1. Certain choices of the perturbation distribution also ensure that the perturbed feasible region has a nonempty interior. Thus, adding randomly drawn perturbations to a stochastic linear program ensures the high-level conditions for uniform Hadamard differentiability hold with probability 1. We then show that this idea can be combined with a functional delta method to construct a uniformly valid confidence set for the true unknown value of the functional or subvector of interest which has exact coverage over a slight expansion of the identified set. The randomness of the perturbation is essential for these theoretical results, which do not hold for non-stochastic perturbations. 

The main tuning parameter in our method determines the support of the perturbations. The theoretical results suggest that even perturbations with extremely small support can correct any uniform Hadamard differentiability failure and allow our approach to obtain the correct asymptotic coverage. However, the current theory is silent on how to set the support of the perturbations to obtain good coverage properties in finite samples. We explore the finite sample impact of the support of the perturbations in simulations. We also compare the method to the methods of \cite{fang2021inference} and \cite{gafarov2021inference} in a variety of simulation examples, and show that the procedure has competitive coverage properties.

\subsection{Related Literature}

A survey of inference methods for partially identified models is provided in \cite{canay2017practical}. The literature has focused on the construction of uniformly valid confidence sets, motivated by the fact that test statistics in moment inequality models have asymptotic distributions that are discontinuous in the moment slackness. In these cases, confidence sets that are only valid pointwise can be highly misleading. 

For subvector inference, \cite{andrews2009validity} and \cite{andrews2010inference} proposed to project confidence sets constructed for the entire parameter vector. While these projection procedures are uniformly valid, they can be highly conservative. Both \cite{romano2008inference} and \cite{bugni2017inference} consider inverting profiled test statistics in order to construct confidence sets for subvectors, and \cite{kaido2019confidence} provide a calibrated projection inference method for functionals of a partially identified parameter. The greatest advantage of our proposed procedure relative to these procedures is in terms of computational tractability, although we restrict attention to linear moment inequalities. Several recent working papers also study inference on functionals or subvectors. \cite{belloni2018subvector} consider the problem of subvector inference in models with many moment inequalities. \cite{andrews2019conditional} consider projection inference for conditional moment inequalities that are linear in nuisance parameters and that also have non-stochastic gradients conditional on some variable. Our approach does not require this special structure, although we require linear moment conditions. \cite{gafarov2021inference} shows how to construct uniformly valid confidence sets in an optimization framework, although relies on a crucial high-level constraint qualification which is not required in our current approach. Finally, \cite{fang2021inference} consider the related problem of testing for the existence of a solution to a linear system with known coefficients. In contrast, our proposed procedure is applicable with both fixed and data-dependent coefficients. However, it is important to keep in mind that our proposed approach ``moves the goalpost'' in the sense that exact coverage is obtained only on an outer set, and the coverage probability is $1$ over the identified set. This a consequence of our unconventional approach based on a global perturbation.

In a relevant paper, \cite{fang2018inference} study the problem of performing inference using a functional delta method for Hadamard differentiable and Hadamard directionally differentiable functionals. They show that full differentiability of the functional of interest is both necessary and sufficient for consistency of the bootstrap when the underlying estimator is asymptotically Gaussian. In the current paper we demonstrate the conditions under which the value functions of a linear program are fully (and uniformly) Hadamard differentiable and show that a perturbation procedure ensures these conditions are satisfied.\footnote{\cite{shapiro1991asymptotic} demonstrates Hadamard directional differentiability of the value function under various assumptions, and also comments on conditions that ensure (full) Hadamard differentiability. } Thus, our theoretical results and our proposed bootstrap procedure do not contradict the results of \cite{fang2018inference}, and allow us to avoid explicitly estimating the Hadamard derivative. Pointwise inference using a functional delta method has also been previously considered in various contexts by \cite{kasy2016partial}, \cite{masten2017inference}, \cite{christensen2019counterfactual}, and \cite{gunsilius2020path}. 

The current paper also has close connections to the literature on support function estimators, including \cite{beresteanu2008asymptotic}, \cite{bontemps2012set}, \cite{kaido2014asymptotically}, and \cite{chandrasekhar2019best}, among others. In particular, \cite{kaido2014asymptotically} propose a pointwise valid procedure that can be used to construct confidence intervals for subvectors in models with convex moment conditions. Their results require both a constraint qualification and uniqueness of the optimal solutions; the ``uniform versions'' of these assumptions are very strong and high-level, which is a main motivation for the current paper. Their inference procedure uses a multiplier bootstrap applied to the score of the support function; in contrast, our approach directly bootstraps the support function after introducing a global perturbation. 

Finally, the current paper is motivated by applications in partial identification with linear moment conditions. There have been a number of applications motivated by the influential papers of \cite{pakes2011imoment,pakes2015moment} that involve linear moment inequalities.  
Other articles relying on linear moment conditions and/or linear programming are now legion in the literature on partial identification and include \cite{honore2006bounds}, \cite{honore2006competing}, \cite{manski2007partial}, \cite{laffers2013note}, \cite{freyberger2015identification}, \cite{demuynck2015bounding}, \cite{mogstad2018using},  \cite{russell2019sharp}, \cite{tebaldi2019nonparametric}, \cite{torgovitsky2019nonparametric}, and \cite{torgovitsky2019partial}, among others. The procedure outlined in this paper is applicable to many of these works. Recently our proposed procedure has been used by \cite{gu2022partial} to construct confidence sets for counterfactual probabilities in a nonseparable binary response model with endogenous regressors, an example with thousands of parameters and constraints.\footnote{In the current paper the number of constraints is assumed to be fixed throughout, and we do not address the case where the number of constraints can grow with the sample size. However, our asymptotic approximations can still be useful in many cases when the number of constraints is large relative to the sample size; for instance, when the constraints depend on the data only through a low dimensional estimated parameter, as is the case in \cite{gu2022partial}.  } 

The remainder of the paper proceeds as follows. Section \ref{section_main_ideas} presents a simplified overview of the method. Section \ref{section_methodology} develops the methodology in detail. Section \ref{section_simulation_evidence} introduces the Monte Carlo exercises performed using our proposed procedure, and Section \ref{section_conclusion} concludes. All appendices are included in the supplementary material. The proofs of all of the main results are provided in Appendix A. A modified version of our proposed procedure that always produces a nonempty confidence interval is presented in Appendix B. Two illustrative examples are presented in Appendix C, and details on our simulation designs are presented in Appendix D, along with some additional robustness exercises.\\

\noindent \textbf{Notation:} The product probability measure is denoted by $\text{Pr}_{P}$. To keep notation clean, we use $\E_{P}[ \,\cdot\,]$ to denote either the expectation with respect to $P$, or the expectation with respect to the product measure. We let $\E_{n}$ denote the expectation with respect to the empirical distribution $\mathbb{P}_{n}$, and we let $\E_{n}^{b}$ denote the expectation with respect to the bootstrap distribution $\mathbb{P}_{n}^{b}$. Issues of measurability are avoided in the main text, but are addressed in the appendix. We use $P_{n} \rsq P$ to denote weak convergence of a sequence of probability measures $\{P_{n}\}_{n=1}^\infty$ to a limiting probability measure $P$.

\section{Overview}\label{section_main_ideas}

The goal of this section is to provide the intuition behind our approach and to provide a roadmap for the remainder of the paper. Our main motivation is to construct a uniformly valid confidence set for the scalar quantity $\psi_{0}:=\E_{P}[\psi(W,\theta_{0})]$, where $W \in \mathcal{W}\subset \mathbb{R}^{d_{w}}$ denotes the relevant finite-dimensional vector of random variables in the model, $\theta_{0} \in \Theta \subset \mathbb{R}^{d_{\theta}}$ is the true but partially identified parameter vector, and $\psi(w,\theta)$ is a function that is linear in $\theta$. We assume throughout that $\Theta$ is a compact polyhedron defined by $k_{\theta}$ linear constraints. We then consider a setting where the identified set $\Theta_{I}(P)$ is defined by $k_{m}$ linear moment inequalities:
\begin{align}
\Theta_{I}(P) := \left\{\theta \in \mathbb{R}^{d_{\theta}} : \E_{P}[m_{j}(W,\theta)]\leq 0,\,\,j=1, \ldots, k \right\},
\end{align}
where $k:= k_{m} + k_{\theta}$. Note that this formulation does not rule out moment equalities, since every moment equality can be written as a combination of two moment inequalities. Under our assumptions presented in the next section, the identified set $\Psi_{I}(P)$ for $\psi_{0}$ is an interval $[\Psi_{I}^{\ell b}(P),\Psi_{I}^{ub}(P)]$ with endpoints determined by the following linear programming problems:
\begin{align}
\Psi_{I}^{\ell b}(P) &:= \min_{\theta \in \mathbb{R}^{d_{\theta}}} \quad \E_{P}[\psi(W,\theta)],&& \text{s.t.} \:\: &\E_{P}[m_{j}(W,\theta)]\leq 0, \qquad j=1, \ldots, k,\label{P_inf}\\
\Psi_{I}^{ub}(P) &:= \max_{\theta \in \mathbb{R}^{d_{\theta}}} \quad \E_{P}[\psi(W,\theta)],&& \text{s.t.} \:\: &\E_{P}[m_{j}(W,\theta)]\leq 0, \qquad j=1, \ldots, k.\label{P_sup}
\end{align}
We are then interested in constructing a random set $C_{n}^{\psi}(1-\alpha)$ that satisfies:
\begin{align}
\liminf_{n \to \infty} \inf_{\{\psi_{0} \in \Psi_{I}(P),\, P \in \mathcal{P}\} } \text{Pr}_{P}(\psi_{0} \in C_{n}^{\psi}(1-\alpha)) \geq 1-\alpha, \label{eq_uniform_CS}  
\end{align}
where $\mathcal{P}$ is some large class of DGPs. We now present some motivating examples which are revisited in our simulation exercises.

\begin{empirical}[Missing Data]\label{example_missing_data}
Consider the canonical missing data example. In this example the researcher observes a sample $\{Y_{i}D_{i}, D_{i}\}_{i=1}^{n}$. For simplicity, suppose that $Y_{i},D_{i} \in \{0,1\}$.  The parameter of interest is the unconditional average of the outcome variable:
\begin{align*}
\E_{P}[\psi(W,\theta)] &:= \psi(\theta) = \sum_{y} \sum_{d} \theta_{yd} \cdot y,
\end{align*}
where $\theta_{yd} := P(Y=y,D=d)$. The constraints imposed by the observed distribution $P(YD=yd,D=d)$ on the latent distribution $\theta_{yd}=P(Y=y,D=d)$ are given by:
\begin{align}
P(YD=0,D=1) &= \theta_{01}, \label{eq_example_missing_data_cons1}\\
P(YD=1,D=1) &= \theta_{11}, \\
P(YD=0, D=0) &= \theta_{00} + \theta_{10}.\label{eq_example_missing_data_cons3}
\end{align}
It is straightforward to see that point identification of $\theta$ occurs only when $P(D=0)=0$. The identified set for our function of interest, $\Psi_{I}(P) = [\Psi_{I}^{\ell b}(P), \Psi_{I}^{ub}(P)]$ can be obtained by solving the problems:
\begin{align}\label{eq_example_missing_data_bounds}
\Psi_{I}^{\ell b}(P) =\min_{\theta \in \Theta_{I}(P)}  \psi(\theta), &&\Psi_{I}^{ub}(P) =\max_{\theta \in \Theta_{I}(P)} \psi(\theta),
\end{align}
where $\Theta_{I}(P)$ is the set of $\theta$ satisfying the constraints \eqref{eq_example_missing_data_cons1} - \eqref{eq_example_missing_data_cons3}. Note that the optimization problems in \eqref{eq_example_missing_data_bounds} are linear programs. 
\end{empirical}

\begin{empirical}[Linear Regression with Interval-Valued Dependent Variable]\label{example_linear_regression}
Consider the example of linear regression with an interval-valued dependent variable.\footnote{We follow closely the exposition in \cite{kaido2019confidence} Appendix C.} The model is given by $Y=X^\top\theta+\varepsilon$, where $X \in \mathbb{R}^{d}$ with $R$ points of support. The value of $Y$ is never observed, although there exists two bounded and observable random variables $Y^{*}$ and $Y_{*}$ such that $P(Y_{*} \leq Y \leq Y^{*})=1$. Denoting the support points of $X$ as $\{x_{1},\ldots,x_{R}\}$, the identified set for $\theta$ is:
\begin{align*}
\Theta_{I}(P) := \{ \theta \in \Theta: \E_{P}[Y_{*}\mid X=x_{r}] - x_{r}^\top\theta \leq 0, \:\: x_{r}^\top\theta - \E_{P}[Y^{*}\mid X=x_{r}] \leq 0, \text{ $r=1,\ldots, R$} \}.
\end{align*}
The objective is to construct a confidence set for the first component $\theta_{1}$ of the parameter vector $\theta$. In our notation, set $\psi(W,\theta) =\theta_{1}$. Under weak conditions, the identified set for the functional $\psi$ is an interval $\Psi_{I}(P) = [\Psi_{I}^{\ell b}(P), \Psi_{I}^{ub}(P)]$ with the endpoints determined by: 
\begin{align}\label{eq_example_linear_regression}
\Psi_{I}^{\ell b}(P) =\min_{\theta \in \Theta_{I}(P)}  \psi(\theta), &&\Psi_{I}^{ub}(P) =\max_{\theta \in \Theta_{I}(P)} \psi(\theta).
\end{align}
The optimization problems in \eqref{eq_example_linear_regression} are linear programs. 
\end{empirical}

\begin{empirical}[Inference on Counterfactual Policies]\label{example_inference_on_counterfactual}
Consider the setting of \cite{kasy2016partial}. Here a treatment variable $D$ determines an outcome $Y$ through the equation: 
\begin{align}
Y = DY_{1} + (1-D)Y_{0},\label{eq_pom}
\end{align}
where $Y_{1}$ and $Y_{0}$ are potential outcomes. Let $X$ denote an observable covariate with finite support. Given a sample $\{(Y_{i},D_{i},X_{i})\}_{i=1}^{n}$, the researcher can construct bounds on the functions $g_{d}(x):= \E_{P}[Y_{d} \mid X=x]$. Let $\mathcal{G}_{d}$ denote the identified set for $g_{d}(\cdot)$. We assume the researcher is then interested in comparing counterfactual treatment assignment policies ``$A$'' and ``$B$.'' Each treatment assignment policy $j \in \{A,B\}$ is characterized by the conditional probability $P(D^{j} =1 \mid X)$. Assume that $D^{A}, D^{B} \independent (Y_{0},Y_{1}) \mid X$. The objective is to construct a confidence set for $\E[Y^{A} - Y^{B}]$, where $Y^{j}$ denotes outcome under policy $j \in \{A,B\}$, determined by the equation: 
\begin{align*}
Y^{j} = D^{j} Y_{1} + (1-D^{j})Y_{0},  
\end{align*}
where $Y_{1}$ and $Y_{0}$ are same potential outcomes from \eqref{eq_pom}. Note that:
\begin{align*}
\E_{P} \left[ Y^{A} - Y^{B}\right]=\E_{P} \left[ (h_{A}(X) - h_{B}(X)) (g_{1}(X) - g_{0}(X))\right],
\end{align*}
where $h_{j}(x) = P(D^{j}=1\mid  X=x)$. Let $\Psi_{I}(P) = [\Psi_{\ell b}(P), \Psi_{u b}(P)]$ denote the identified set for $\E_{P} \left[ Y^{A} - Y^{B}\right]$. The endpoints of $\Psi_{I}^{AB}(P)$ are given by:
\begin{align}
&\Psi_{\ell b}(P) = \min_{(g_{0},g_{1}) \in \mathcal{G}_{0} \times \mathcal{G}_{1}} \sum_{x \in \mathcal{X}}(h_{A}(x) - h_{B}(x)) (g_{1}(x) - g_{0}(x)) P(X=x),\label{eq_kasy2_lb}\\
&\Psi_{u b}(P) = \max_{(g_{0},g_{1}) \in \mathcal{G}_{0} \times \mathcal{G}_{1}} \sum_{x \in \mathcal{X}}(h_{A}(x) - h_{B}(x)) (g_{1}(x) - g_{0}(x))  P(X=x). \label{eq_kasy2_ub}
\end{align}
In this example, note that the partially identified parameter is $\theta = (g_{0},g_{1})$, a $2 \times |\mathcal{X}|$ vector. The identified set $\Theta_{I}(P) = \mathcal{G}_{0} \times \mathcal{G}_{1}$ can be characterized by linear constraints, so the problems \eqref{eq_kasy2_lb} and \eqref{eq_kasy2_ub} are linear programs.
\end{empirical}
Our method of inference for $\psi_{0}$ relies on an approximation to the distribution of the sample analog value functions $\Psi_{I}^{\ell b}(\P_{n})$ and $\Psi_{I}^{ub}(\P_{n})$. Focusing on the lower bound value function $\Psi_{I}^{\ell b}(\P_{n})$, note that:
\begin{align}
\sqrt{n}(\Psi_{I}^{\ell b}(\P_{n}) - \Psi_{I}^{\ell b}(P_{n})) = \frac{\Psi_{I}^{\ell b}(P_{n} + t_{n} h_{n} ) - \Psi_{I}^{\ell b}(P_{n})}{t_{n}},\label{eq_uhd_derivative_lb}
\end{align}
for any sequence $P_{n}$, where $t_{n}:= n^{-1/2}$ and $h_{n}:= t_{n}^{-1} (\P_{n} - P_{n})$. Under some conditions, if $P_{n}$ converges (in a sense to be made precise) to some $P$, then \eqref{eq_uhd_derivative_lb} tends to a limit called a uniform Hadamard derivative.\footnote{In the proofs, we show that it is without loss of generality to assume $P_{n}$ converges weakly to some $P$, which we denote by $P_{n}\rsq P$. In particular, under our assumptions our class of DGPs $\mathcal{P}$ is tight and closed in the weak* topology. Combined with Prokhorov's Theorem (e.g. \cite{billingsley1968convergence} Theorem 6.1), every sequence $\{P_{n}\}_{n=1}^{\infty}$ in $\mathcal{P}$ admits a weakly convergent subsequence converging to some $P \in \mathcal{P}$.} This is demonstrated in Theorem \ref{thm_uhd} in the next section. Functions that are differentiable in this sense are said to be uniformly Hadamard differentiable (UHD), and are amenable to an application of the delta method. We show that the value function $\Psi_{I}^{\ell b}(\,\cdot\,)$ is UHD at $P$ if:
\begin{enumerate}[label=(\roman*)]
 	\item The identified set $\Theta_{I}(P)$ has nonempty interior at $P$.
 	\item The Lagrange multipliers for the program \eqref{P_inf} exist and are unique at $P$.
 	\item The optimal solutions in the program \eqref{P_inf} are unique at $P$.
\end{enumerate}
Condition (ii) implies that there cannot be more than $d_{\theta}$ binding constraints at any solution to the linear programs \eqref{P_inf} and \eqref{P_sup}, and condition (iii) rules out cases when the solutions to \eqref{P_inf} and \eqref{P_sup} occur at a ``flat face'' of the identified set.\footnote{Similar conditions are encountered in \cite{hirano2012impossibility} and \cite{kaido2014asymptotically}.} All three conditions are high-level.   

To make progress, we turn to the literature on \textit{genericity} in linear programming problems. Intuitively, a generic property of a problem is a property that holds in ``almost all'' instances of the problem. In particular, consider the following  linear programs, which are revised versions of \eqref{P_inf}:
\begin{align}
\Psi_{I,-}^{\ell b}(P,\xi) &:= \min_{\theta \in \mathbb{R}^{d_{\theta}}} \quad \E_{P}[\psi(W,\theta)] - \langle \nu, \theta \rangle\quad \text{s.t.} \quad \E_P[m_{j}(W,\theta)]\leq \varepsilon_{j}, \qquad j=1, \ldots, k,\label{eq_P_inf_xi_minus}\\
\Psi_{I,+}^{\ell b}(P,\xi) &:= \min_{\theta \in \mathbb{R}^{d_{\theta}}} \quad \E_{P}[\psi(W,\theta)]+ \langle \nu, \theta \rangle\quad  \text{s.t.} \quad \E_P[m_{j}(W,\theta)]\leq \varepsilon_{j}, \qquad j=1, \ldots, k.\label{eq_P_inf_xi_plus}
\end{align}
Here $\xi :=(\nu,\varepsilon) \in \mathbb{R}^{d_{\theta}}\times \mathbb{R}_{+}^{k}$ is a draw from some known distribution $P_{\xi}$. We call these random draws \textit{perturbations} and call the linear programs \eqref{eq_P_inf_xi_minus} and \eqref{eq_P_inf_xi_plus} \textit{perturbed linear programs}. In practice the perturbation $\xi$ can be drawn from any continuous distribution with convex support, although it must be independent of $W$. It is possible to show:
\begin{align}
\min\{ \Psi_{I,-}^{\ell b}(P,\xi), \Psi_{I,+}^{\ell b}(P,\xi)\} < \Psi_{I}^{\ell b}(P) \leq \psi_{0} \leq \Psi_{I}^{\ell b}(P) < \max\{ \Psi_{I,-}^{u b}(P,\xi), \Psi_{I,+}^{u b}(P,\xi)\}, \label{eq_perturbed_inequality}
\end{align}
$P_{\xi}-$a.s, where $\Psi_{I,-}^{u b}(P,\xi)$ and $\Psi_{I,+}^{u b}(P,\xi)$ are the analogous upper bound perturbed value functions. That is, the perturbed programs can be used to outer-bound the identified set for $\psi_{0}$. 

The results of \cite{spingarn1979generic} imply that conditions (i), (ii) and (iii) hold for the perturbed linear programs $P_{\xi}-$a.s. at any $P \in \mathcal{P}$. This property implies that the value functions in the perturbed linear programs are UHD $P_{\xi}-$a.s.\footnote{The perturbation idea here appears to be similar the use of data-jittering in \cite{chandrasekhar2019best}. \cite{bontemps2012set} also study the generic differentiability of support functions.} 
Thus, for $P_{\xi}-$almost all perturbations $\xi$, we can use the functional delta method to approximate the distributions of the value functions $\Psi_{I,-}^{\ell b}(P,\xi)$, $\Psi_{I,+}^{\ell b}(P,\xi)$, $\Psi_{I,-}^{u b}(P,\xi)$, and $\Psi_{I,+}^{u b}(P,\xi)$, and can then use the inequality from \eqref{eq_perturbed_inequality} to construct a uniformly valid confidence set for $\psi_{0}$. This idea is fully developed in Section \ref{subsection_confidence_sets}. Intuitively, the perturbation eliminates all over-identified solutions of the linear program, and ensures that a single basic solution of the perturbed linear program is picked out asymptotically.\footnote{We thank an anonymous referee for pointing out this perspective.} Inference is then done around this picked-out solution. The perturbation is designed in a way to ensure that using this solution for inference introduces at most a conservative distortion. We then show that a simple nonparametric bootstrap can be used to approximate the asymptotic distribution of the perturbed value functions. In the end, a confidence set $C_{n}^{\psi}(1-\alpha,\xi)$ for $\psi_{0}$ can be constructed as follows:
\begin{enumerate}[label=Step \arabic*:,leftmargin=*]
\item Choose small $\bar{\varepsilon}>0$, and draw $\xi$ from the uniform distribution on $\Xi:=[0,\bar{\varepsilon}]^{d_{\theta}+k}$.\footnote{ In our simulations (Section \ref{section_simulation_evidence}) $\xi$ is drawn from the uniform distribution on $[0,10^{-3}]^{d_{\theta}+k}$.} 
\item Resample $\{W_{i}^{b}\}_{i=1}^{n}$ i.i.d. with replacement with equal probability from $\{W_{i}\}_{i=1}^{n}$, compute the bootstrap objective function and moment conditions:
\begin{align*}
\E_{n}^{b}[\psi(W,\theta)]:= \frac{1}{n} \sum_{i=1}^{n} \psi(W_{i}^{b},\theta), &&\E_{n}^{b}[m_{j}(W,\theta)]:= \frac{1}{n} \sum_{i=1}^{n}  m_{j}(W_{i}^{b},\theta), \text{ for  $j=1,\ldots,k$,}
\end{align*}
and solve the linear programs:\mathtoolsset{showonlyrefs=false}
\begin{align}
\Psi_{I,-}^{\ell b}(\P_{n}^{b},\xi) := \min_{\theta \in \mathbb{R}^{d_{\theta}}}\,\, &\E_{n}^{b}[\psi(W,\theta)] - \langle \nu, \theta \rangle\quad \text{s.t.} \quad \E_{n}^{b}[m_{j}(W,\theta)]\leq \varepsilon_{j}, \,\,j=1, \ldots, k,\label{eq_P_inf_perturb_bootstrap_minus}\\
\Psi_{I,+}^{\ell b}(\P_{n}^{b},\xi) := \min_{\theta \in \mathbb{R}^{d_{\theta}}}\,\,&\E_{n}^{b}[\psi(W,\theta)]+ \langle \nu, \theta \rangle\quad\text{s.t.} \quad\E_{n}^{b}[m_{j}(W,\theta)]\leq \varepsilon_{j},\,\,j=1, \ldots, k,\label{eq_P_inf_perturb_bootstrap_plus}\\
\Psi_{I,-}^{ub}(\P_{n}^{b},\xi) := \max_{\theta \in \mathbb{R}^{d_{\theta}}}\,\, &\E_{n}^{b}[\psi(W,\theta)]- \langle \nu, \theta \rangle\quad\text{s.t.}\quad\E_{n}^{b}[m_{j}(W,\theta)]\leq \varepsilon_{j},\,\, j=1, \ldots, k,\label{eq_P_sup_perturb_bootstrap_minus}\\
\Psi_{I,+}^{ub}(\P_{n}^{b},\xi) := \max_{\theta \in \mathbb{R}^{d_{\theta}}}\,\,&\E_{n}^{b}[\psi(W,\theta)]+ \langle \nu, \theta \rangle\quad\text{s.t.}\quad\E_{n}^{b}[m_{j}(W,\theta)]\leq \varepsilon_{j}, \,\,j=1, \ldots, k,\label{eq_P_sup_perturb_bootstrap_plus}
\end{align}
where $\xi=(\nu,\varepsilon)$ is the perturbation from Step 1. 
\item Repeat Step 2 for $b=1,\ldots,\overline{B},$ where $\overline{B}$ is some large integer.
\item Solve the linear programs:
\begin{align}
\Psi_{I,-}^{\ell b}(\P_{n},\xi) := \min_{\theta \in \mathbb{R}^{d_{\theta}}} \,\, &\E_{n}[\psi(W,\theta)] - \langle \nu, \theta \rangle\quad\text{s.t.}\quad\E_{n}[m_{j}(W,\theta)]\leq \varepsilon_{j},  \,\, j=1, \ldots, k,\label{eq_P_inf_perturb_empirical_minus}\\
\Psi_{I,+}^{\ell b}(\P_{n},\xi) := \min_{\theta \in \mathbb{R}^{d_{\theta}}}\,\, &\E_{n}[\psi(W,\theta)]+ \langle \nu, \theta \rangle\quad\text{s.t.}\quad\E_{n}[m_{j}(W,\theta)]\leq \varepsilon_{j},  \,\,j=1, \ldots, k,\label{eq_P_inf_perturb_empirical_plus}\\
\Psi_{I,-}^{ub}(\P_{n},\xi) := \max_{\theta \in \mathbb{R}^{d_{\theta}}}\,\, &\E_{n}[\psi(W,\theta)]- \langle \nu, \theta \rangle\quad\text{s.t.} \quad\E_{n}[m_{j}(W,\theta)]\leq \varepsilon_{j},  \,\,j=1, \ldots, k,\label{eq_P_sup_perturb_empirical_minus}\\
\Psi_{I,+}^{ub}(\P_{n},\xi) := \max_{\theta \in \mathbb{R}^{d_{\theta}}}\,\, &\E_{n}[\psi(W,\theta)]+ \langle \nu, \theta \rangle\quad\text{s.t.}\quad\E_{n}[m_{j}(W,\theta)]\leq \varepsilon_{j}, \,\, j=1, \ldots, k.\label{eq_P_sup_perturb_empirical_plus}
\end{align}
\item Set:
\begin{align}
\gamma(\alpha):=(\alpha)\mathbbm{1}\{\Delta(\P_{n},\xi) > b_{n}\} + (\alpha/2)\mathbbm{1}\{\Delta(\P_{n},\xi) \leq b_{n}\},\label{eq_gamma_correction}
\end{align}
for some sequence $b_{n}\downarrow 0$, where:
\begin{align*}
\Delta(\P_{n},\xi):= \max\{\Psi_{I,+}^{ub}(\P_{n},\xi), \Psi_{I,-}^{ub}(\P_{n},\xi)\} - \min\{\Psi_{I,+}^{\ell b}(\P_{n},\xi), \Psi_{I,-}^{\ell b}(\P_{n},\xi)\}.
\end{align*}
Compute the $(1-\gamma(\alpha))$\textsuperscript{th} quantiles $\Psih_{\alpha,\xi,-}^{\ell b}$, $\Psih_{\alpha,\xi,+}^{\ell b}$, $\Psih_{\alpha,\xi,-}^{u b}$ and $\Psih_{\alpha,\xi,+}^{u b}$ of the bootstrap distributions of:\label{eq_step_alpha}
\begin{align}
\sqrt{n}&(\Psi_{I,-}^{\ell b}(\mathbb{P}_{n}^{b},\xi) - \Psi_{I,-}^{\ell b}(\P_{n},\xi)),\label{eq_bootstrap_quantity1}\\
\sqrt{n}&(\Psi_{I,+}^{\ell b}(\mathbb{P}_{n}^{b},\xi) - \Psi_{I,+}^{\ell b}(\P_{n},\xi)),\label{eq_bootstrap_quantity2}\\
-\sqrt{n}&(\Psi_{I,-}^{u b}(\mathbb{P}_{n}^{b},\xi) - \Psi_{I,-}^{u b}(\P_{n},\xi)),\label{eq_bootstrap_quantity3}\\
-\sqrt{n}&(\Psi_{I,+}^{u b}(\mathbb{P}_{n}^{b},\xi) - \Psi_{I,+}^{u b}(\P_{n},\xi)).\label{eq_bootstrap_quantity4}
\end{align}

\item Construct the confidence set $C_{n}^{\psi}(1-\alpha,\xi)$ as follows:
\begin{align}
C_{n}^{\psi}(1-\alpha,\xi) := \left[C_{\ell b, n}^{\psi}(1-\alpha,\xi),\,\, C_{u b, n}^{\psi}(1-\alpha,\xi)\right],\label{CS_proposal}
\end{align}
where:\mathtoolsset{showonlyrefs=true}
\begin{align}
C_{\ell b, n}^{\psi}(1-\alpha,\xi)&=\min\{\Psi_{I,-}^{\ell b}(\mathbb{P}_{n},\xi),\Psi_{I,+}^{\ell b}(\mathbb{P}_{n},\xi)\} - \frac{\max\left\{\Psih_{\alpha,\xi,-}^{\ell b},\Psih_{\alpha,\xi,+}^{\ell b}\right\}}{\sqrt{n}},\\
C_{u b, n}^{\psi}(1-\alpha,\xi)&=\max\{\Psi_{I,-}^{u b}(\mathbb{P}_{n},\xi),\Psi_{I,+}^{u b}(\mathbb{P}_{n},\xi)\} + \frac{\max\left\{\Psih_{\alpha,\xi,-}^{u b},\Psih_{\alpha,\xi,+}^{u b}\right\}}{\sqrt{n}}.
\end{align}
\end{enumerate}
\begin{remark}
The procedure presumes that the linear programs in Steps 2 and 4 have nonempty feasible regions. When this is not the case, the procedure can instead be applied to relaxed versions of these linear programs at the expense of producing a larger confidence interval. A detailed procedure for computing an always nonempty confidence set is presented in Appendix B. 
\end{remark}
\begin{remark}
Asymptotically, our method is valid taking the $(1-\alpha)^{\text{th}}$ quantile in Step 5 of the procedure above, although we find that using the $(1-\gamma(\alpha))^{th}$ improves finite sample performance when the set $\Psi_{I}(\P_{n},\xi)$ is small relative to sampling uncertainty. The impact of this quantity is investigated in Appendix D.5. 
\end{remark}
To understand the coverage properties of this confidence set, define:
\begin{align}
\overline{\Psi}_{I}(P) := \bigcup_{\xi \in \Xi} \Psi_{I}(P,\xi), \label{eq_maximally_relaxed}
\end{align}
where:
\begin{align*}
\Psi_{I}(P,\xi) :=[\Psi_{I}^{\ell b}(P,\xi), \Psi_{I}^{u b}(P,\xi)],
\end{align*}
\begin{align*}
\Psi_{I}^{\ell b}(P,\xi):=\min\{\Psi_{I,-}^{\ell b}(P,\xi),\Psi_{I,+}^{\ell b}(P,\xi)\},&&
\Psi_{I}^{u b}(P,\xi):=\max\{\Psi_{I,-}^{u b}(P,\xi),\Psi_{I,+}^{u b}(P,\xi)\}.
\end{align*}
Intuitively, the set $\Psi_{I}(P,\xi)$ represents a relaxed version of the identified set $\Psi_{I}(P)$, where the relaxation is determined by the perturbation $\xi$. The set $\overline{\Psi}_{I}(P)$ then contains all possible sets $\Psi_{I}(P,\xi)$, and so represents a ``maximally relaxed'' version of the identified set for a given value of $\bar{\varepsilon}$. Under weak assumptions, Theorem \ref{corollary_uniformity} states that the confidence set $C_{n}^{\psi}(1-\alpha,\xi)$ satisfies:
\begin{align}
\liminf_{n \to \infty} \inf_{\{ (\psi,P): \:\: \psi \in \Psi_{I}(P), \:\:P \in \mathcal{P} \}} (\text{Pr}_{P}\times P_{\xi})(\psi \in C_{n}^{\psi}(1-\alpha,\xi)) =1.\label{eq_main_theorem_22}
\end{align}
That is, our confidence set contains the true value $\psi_{0}$ with probability $1$ asymptotically over repeated draws from both the sampling distribution $\text{Pr}_{P}$ and the perturbation distribution $P_{\xi}$. Any tests based on inverting our confidence set will also have zero power for local alternatives drifting to a point in the identified set. However, under some additional assumptions, Theorem \ref{corollary_uniformity} also shows that:
\begin{align}
\liminf_{n \to \infty} \inf_{\{ (\psi,P): \:\: \psi \in \widetilde{\Psi}_{\alpha}(P), \:\:P \in \mathcal{P} \}} (\text{Pr}_{P}\times P_{\xi})(\psi \in C_{n}^{\psi}(1-\alpha,\xi)) = 1-\alpha,
\end{align}
where $\widetilde{\Psi}_{\alpha}(P)$ is a set satisfying $\Psi_{I}(P) \subset \widetilde{\Psi}_{\alpha}(P) \subset \overline{\Psi}_{I}(P)$. In other words, the confidence set $C_{n}^{\psi}(1-\alpha,\xi)$ obtains exact uniform coverage over a set $\widetilde{\Psi}_{\alpha}(P)$ that lies between the identified set $\Psi_{I}(P)$ and the maximally relaxed version of the identified set from \eqref{eq_maximally_relaxed}. When the support of the perturbations is small, the identified set $\Psi_{I}(P)$ and the maximally relaxed set $\overline{\Psi}_{I}(P)$ are typically very close, implying that the confidence set $C_{n}^{\psi}(1-\alpha,\xi)$ has exact uniform coverage on a slight expansion of the identified set. In this sense the proposed approach moves the objective, obtaining exact coverage on an outer set in exchange for an inference procedure that is computationally and conceptually simple. 

The randomness of the perturbation is essential for this theoretical result, which does not hold for non-stochastic perturbations. In particular, there can exist (at most) a $P_{\xi}-$null set $\Xi_{0} \subset \Xi$ for which the uniform coverage probability is below the nominal level. If a non-stochastic perturbation is used, and if this perturbation belongs to $\Xi_{0}$, then the coverage guarantee will fail. A randomly drawn perturbation lies outside of this problematic null set $P_{\xi}-$a.s., in which case our coverage results are guaranteed.

Our theoretical results require a positive and constant $\bar{\varepsilon}$ to construct the support of the perturbations $\Xi$ in Step 1 of the procedure above. The value of $\bar{\varepsilon}$ represents the upper bound on both the objective function and constraint perturbations. The uniform asymptotic validity of the proposed confidence set is not affected as long as $\bar{\varepsilon} > 0$, so in theory it may be chosen as small as desired. The fact that $\bar{\varepsilon}$ is not scale invariant also does not affect the asymptotic theory in the paper. However, the current theoretical results do not hold if $\bar{\varepsilon} = 0$ or if $\bar{\varepsilon} \downarrow 0$ as $n \to \infty$. In addition, the asymptotic theory is silent on the impact of $\bar{\varepsilon}$ in finite samples. We discuss some of the finite sample tradeoffs involved in picking $\bar{\varepsilon}$ in Section \ref{section_ebar_choice} in a simulation example.

In the end, the computational advantage of the proposed procedure arises from the fact that we do not construct our confidence set by inverting hypothesis tests. Instead, our proposed procedure solves $4(B+1)$ linear programs, each of which can be solved in a fraction of a second even with thousands of parameters and constraints. The result is a confidence set $C_{n}^{\psi}(1-\alpha,\xi)$ that is valid under weak assumptions, can be computed in seconds, and that has exact uniform asymptotic coverage on a small expansion of the identified set.

\section{Methodology}\label{section_methodology}

\subsection{Main Assumptions}\label{subsection_main_assumptions}

In this section we introduce the assumptions required to prove the uniform asymptotic validity of our confidence set. 

\begin{assumption}\label{assump_continuity} For $\mathcal{W} \subset \mathbb{R}^{d_{w}}$, we have: (i) the functional of interest $\psi:\mathcal{W} \times \mathbb{R}^{d_{\theta}} \to \mathbb{R}$ is linear in $\theta$, and is continuous in $w \in \mathcal{W}$; (ii) the functions $m_{j}: \mathcal{W} \times \mathbb{R}^{d_{\theta}} \to \mathbb{R}$ are linear in $\theta$ and continuous in $w \in \mathcal{W}$ for $j=1,\ldots,k_{m}$.
\end{assumption}
Note when $\mathcal{W}$ is finite or countable, we can equip $\mathcal{W}$ with the discrete topology, in which case every function on $\mathcal{W}$ becomes both continuous and Borel measurable, and the continuity requirement in Assumption \ref{assump_continuity} is vacuous. In the next assumption, we take $\Theta \subset \mathbb{R}^{d_{\theta}}$ and $\mathcal{P} \subset \mathscr{P}$ with $\mathscr{P}$ the collection of all probability measures on $\mathcal{W}$.
\begin{assumption}\label{assump_regularity}
The parameter space $(\Theta, \mathcal{P})$ satisfies the following conditions:
\begin{enumerate}[label=(\roman*)]
	\item $\Theta:= \{\theta \in \mathbb{R}^{d_{\theta}} : A \theta \leq b \}$ is a compact polyhedron characterized by $k_{\theta}$ linear inequality constraints for a known $k_{\theta} \times d_{\theta}$ matrix $A$ and a known $k_{\theta}\times 1$ vector $b$. These inequality constraints are written as linear moment functions $m_{k_{m}+1}(w,\theta),\ldots,m_{k}(w,\theta)$, where $k:=k_{m}+k_{\theta}$. 
	\item For every $P \in \mathcal{P}\subset \mathscr{P}$ there exists some $\theta \in \Theta$ such that $\E_{P}[m_{j}(W,\theta)]\leq 0$, for $j=1, \ldots, k$. 
	\item In a sample $\{W_{i}\}_{i=1}^{n}$, $W_{i} \in \mathcal{W}\subset \mathbb{R}^{d_{w}}$ are independent and identically distributed according to some $P\in \mathcal{P}$.
	\item For some finite $\overline{\varepsilon}>0$, let:\mathtoolsset{showonlyrefs=false}
	\begin{align}
	\mathcal{F} &:= \left\{\left(\psi(\,\cdot\,,\theta), m_{1}(\,\cdot\,,\theta),\ldots,m_{k}(\,\cdot\,,\theta)\right)^\top : \mathcal{W} \to \mathbb{R}^{k+1} \mid \theta\in \overline{\Theta}\right\},\label{eq_function_class} \\ 
	\overline{\Theta}&:=\{\theta \in \mathbb{R}^{d_{\theta}} : A \theta \leq b + \overline{\varepsilon} \}.
	\end{align}\mathtoolsset{showonlyrefs=true}
	Then there exists an element-wise measurable envelope function $F: \mathcal{W} \to \mathbb{R}$ for the class $\mathcal{F}$ that is bounded on $\mathcal{W}$.
	\item There exists a $B<\infty$ and an $a>0$ such that $\sup_{P\in \mathcal{P}} \E_{P}[||W||^{2+a}]\leq B$.
	\item There exists some constant $C<\infty$ such that:
	\begin{align}
	\sup_{P\in \mathcal{P}}\max\{\E_{P}||\nabla_{\theta} \psi(W,\theta)||^{2}, \E_{P}||\nabla_{\theta} m_{1}(W,\theta)||^{2}, \ldots, \E_{P}||\nabla_{\theta} m_{k}(W,\theta)||^{2} \} \leq C. 
	\end{align}
\end{enumerate}
\end{assumption}

The compactness of $\Theta$ in part (i) of Assumption \ref{assump_regularity} is standard, and ensures that the identified set for our functional of interest is a closed interval.  Part (i) also clarifies that all parameter space constraints are written as moment functions $m_{k_{m}+1}(w,\theta),\ldots,m_{k}(w,\theta)$, where $k:=k_{m}+k_{\theta}$, in order to simplify notation throughout. Part (ii) restricts $\mathcal{P}$ to be a class of DGPs satisfying a set of moment inequalities. Note that we do not rule out moment equalities.\footnote{In particular, $x=0$ can be written as $x \leq 0$ and $-x \leq 0$.} Part (iv) requires the existence of a bounded envelope function, and part (v) is a standard uniform moment condition.\footnote{The boundedness of the envelope $F$ is slightly stronger than the usual moment condition found in the literature, which typically has the form: $\sup_{P \in \mathcal{P}} \E_{P}|F(W)|^{2+a}  < \infty$ for some $a>0$.  However, in practice the existence of a bounded measurable envelope function is more easily verifiable.} Assumptions similar to part (vi) are also commonly found in the existing literature; for instance, \cite{bugni2017inference} A.3(c) (coupled with compactness of $\Theta$) or \cite{kaido2019constraint} Assumption E.4(ii). 

\subsection{Differentiability of the Value Functions}\label{subsection_differentiability}

In this section we provide conditions under which the value function of a linear program is UHD. A formal definition of a UHD functional is presented in Definition A.6 in the Supplementary Material. Let $I(\theta,P)$ be the set indexing the binding moment inequalities at $\theta$ for some probability measure $P \in \mathcal{P}$:
\begin{align}
    I(\theta,P):= \{ j \in \{1,\ldots,k\} : \E_{P}[m_{j}(W,\theta)] = 0  \}.
\end{align}
Furthermore, let $\theta_{\ell b}^{*}(P)$ and $\theta_{u b}^{*}(P)$ be any optimal solutions to the problems \eqref{P_inf} and \eqref{P_sup}, and let $G(\theta,P)$ be the matrix formed by vertically stacking the row vectors $\{\nabla_{\theta} \E_{P}[m_{j}(W,\theta)] \}_{j \in I(\theta,P)}$. 
\begin{definition}[Conditions (NE), (CQ) and (US)]\label{definition_neq_cq_us}
Consider the optimization problems in \eqref{P_inf} and \eqref{P_sup}. We say that $P$ satisfies condition (NE) (for ``non-emptiness'') if $\text{int}(\Theta_{I}(P))\neq \emptyset$. We say that $P$ satisfies condition (CQ) (for ``constraint qualification'') if:
\begin{align}
\min \left\{ eig\left(G(\theta_{\ell b}^{*}(P),P)G(\theta_{\ell b}^{*}(P),P)^\top \right), eig\left(G(\theta_{u b}^{*}(P),P)G(\theta_{u b}^{*}(P),P)^\top \right)\right\} >0,
\end{align}
where $eig(A)$ denotes the minimum eigenvalue of the matrix $A$. Finally, we say that $P$ satisfies condition (US) (for ``unique solutions'') if the optimal solutions $\theta_{\ell b}^{*}(P)$ and $\theta_{u b}^{*}(P)$ exist and are unique.
\end{definition}

Our first result shows that if $P \in \mathcal{P}$ satisfies conditions (NE), (CQ) and (US), then the value functions are UHD at $P$. The tangent cone $\mathcal{T}_{P}(\mathcal{F})$ in the following result is defined in Appendix A.2 of the Supplementary Material. 
\begin{theorem}\label{thm_uhd}
Suppose Assumptions \ref{assump_continuity} and \ref{assump_regularity} hold, and consider the maps $\Psi_{I}^{\ell b},\Psi_{I}^{u b}: \mathscr{P}\to \mathbb{R}$ defined by the programs \eqref{P_inf} and \eqref{P_sup}. Suppose $P\in\mathcal{P}$ satisfies conditions (NE), (CQ) and (US) from Definition \ref{definition_neq_cq_us}. Then $P \mapsto \Psi_{I}^{\ell b}(P) ,\Psi_{I}^{u b}(P)$ are uniformly Hadamard differentiable at $P$ tangential to $\mathcal{T}_{P}(\mathcal{F})$. 
\end{theorem}
Theorem \ref{thm_uhd} represents one of the key results in this paper, and shows that under our assumptions and conditions (NE), (CQ) and (US), the value functions of a linear program are UHD. Condition (NE) is required to ensure that the uniform Hadamard derivative is well-defined. This may not be the case if the identified set at $P$ consists of a single point, since in such cases it is possible to construct a sequence $P_{n} \in \mathscr{P}$ converging to $P \in \mathcal{P}$ such that $\Theta_{I}(P_{n})$ is empty for all $n$. However, condition (NE) rules out both moment equalities and point identification. In this sense, condition (NE) is quite restrictive. In the next subsection we show how the condition can be relaxed to allow for both moment equalities and point identification. Condition (CQ) implies the \textit{linear independence constraint qualification} (LICQ) used in the study of optimality conditions for convex programs.\footnote{\cite{wachsmuth2013licq} shows that the LICQ is the weakest constraint qualification under which the Lagrange multipliers are guaranteed to be unique. Constraint qualifications in various forms have appeared throughout the recent history of partial identification (e.g. \cite{beresteanu2008asymptotic}, \cite{pakes2011imoment} \cite{kaido2014asymptotically}, \cite{freyberger2015identification}, \cite{kaido2019confidence}, \cite{gafarov2018delta}, and \cite{gafarov2021inference}). We refer to the recent paper of \cite{kaido2019constraint} for a full comparison of the constraint qualifications used in the partial identification literature.} There are some cases where it may be easy to directly verify that condition (CQ) is satisfied; for example, when the gradients of the moment functions and objective functions are known. 
However, with data-dependent gradients condition (CQ) is a high-level condition. Condition (US) then imposes uniqueness of optimal solutions separately. If the problems \eqref{P_inf} and \eqref{P_sup} admit multiple solutions, the \textit{limiting optimal solutions} (i.e. over a sequence $P_{n} \to P$ in $\mathcal{P}$) can differ from the \textit{optimal solutions at the limit} (i.e. at $P \in \mathcal{P}$).\footnote{This is related to the Theorem of the Maximum, which guarantees only that the solution correspondence is upper hemicontinuous, but not lower hemicontinuous (and thus, not continuous).} In this case it is possible to show that the value functions $\Psi_{I}^{\ell b}(\cdot)$ and $\Psi_{I}^{u b}(\cdot)$ may not even be Hadamard differentiable, let alone UHD. A similar intuition applies to the Lagrange multipliers. Examples showing the failure of Hadamard differentiability when either (US) or (CQ) fails to hold are presented in Appendix C. 

\begin{figure}[!t]
\centering
\begin{subfigure}[!t]{0.42\textwidth}
\centering
\includegraphics[width=\textwidth]{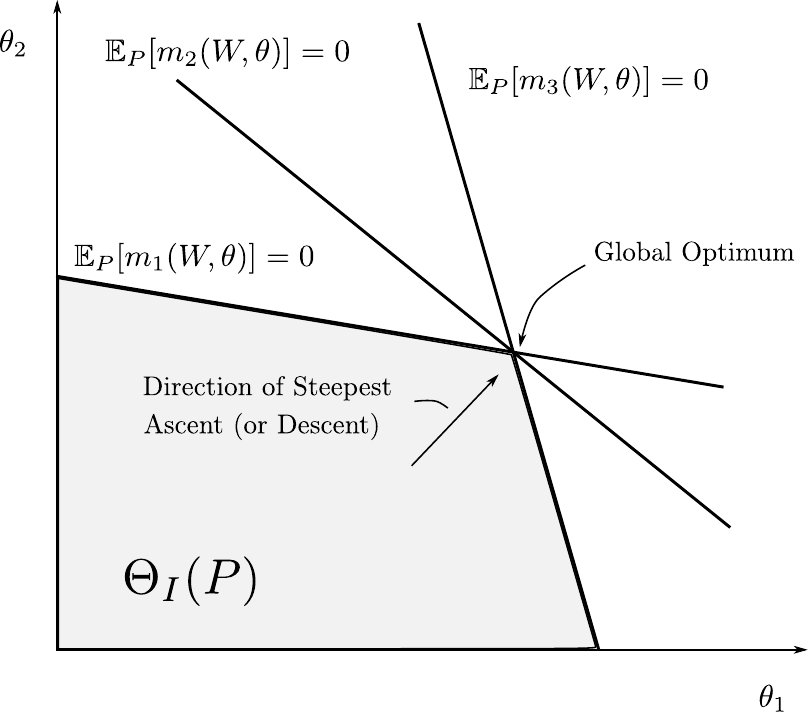}
\caption{This figure illustrates the failure of condition (CQ). In particular, three constraints bind at the global optimum in $d_{\theta}=2$ dimensional space, implying that the gradients of the binding constraints cannot be linearly independent.} \label{fig_cq_failure}
\end{subfigure}
\hfill
\begin{subfigure}[!t]{0.42\textwidth}
\centering
\includegraphics[width=\textwidth]{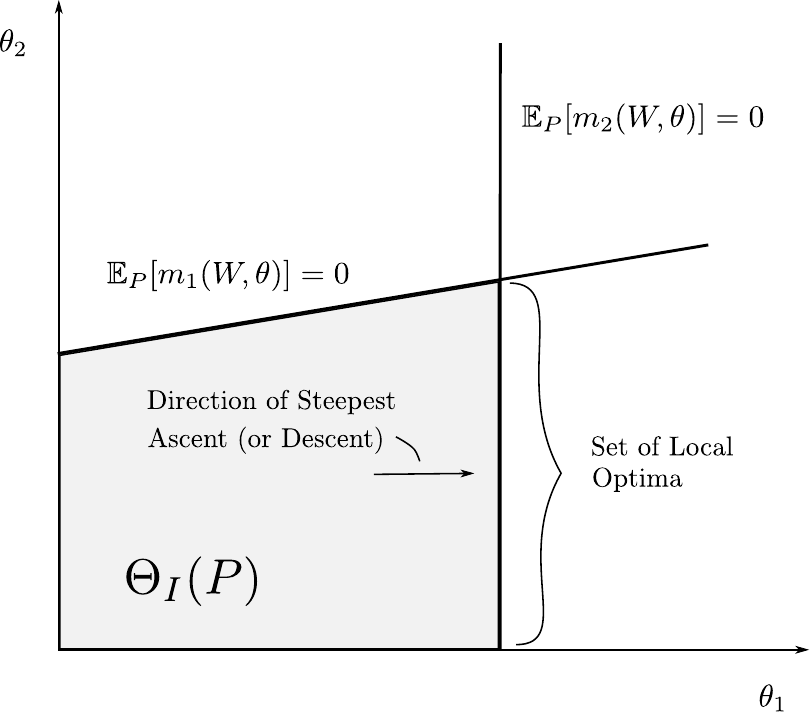}
\caption{This figure illustrates the failure of condition (US). In particular, the objective function is maximized (or minimized) on a ``flat face'' of the feasible region, implying many pairs $(\theta_{1},\theta_{2})$ are optimal.\\} \label{fig_us_failure}
\end{subfigure}
\caption{An illustration of failures of conditions (CQ) and (US).}\label{fig_failure_cq_us}
\end{figure}

Failures of conditions (CQ) and (US) are illustrated in Figure \ref{fig_failure_cq_us}. Figure \ref{fig_failure_cq_us}(a) shows that condition (CQ) fails when the number of binding constraints exceeds the dimension of the parameter space, in which case it is impossible for the gradients of the binding constraints to be linearly independent. Figure \ref{fig_failure_cq_us}(b) also shows that condition (US) fails when the gradient of the objective function is parallel to the gradient of one of the constraints, so that the optimum is obtained along a ``flat face'' of the feasible region. 

Unfortunately, all three conditions (NE), (CQ) and (US) are high-level, although in the next section we show that introducing a simple random perturbation ensures that conditions (NE), (CQ) and (US) are satisfied almost surely.

\subsection{The Perturbation Approach}\label{subsection_perturbation}

Let $\xi =(\nu,\varepsilon)\in \mathbb{R}^{d_{\theta}}\times \mathbb{R}_{+}^{k}$ be a random vector. We impose the following assumption on the distribution and support of $\xi$.   
\begin{assumption}\label{assumption_perturbation}
The random perturbation $\xi$ is independent of $\{W_{i}\}_{i=1}^{n}$ and has distribution $P_{\xi}$ which is absolutely continuous with respect to the Lebesgue measure and has support on some convex set $\Xi \subset \mathbb{R}^{d_{\theta}}\times [0,\overline{\varepsilon}]^{k}$ containing an open set for the same $\overline{\varepsilon}>0$ from Assumption \ref{assump_regularity}.
\end{assumption}
In practice the distribution $P_{\xi}$ is chosen by the researcher, and any distribution can be chosen so long as it satisfies Assumption \ref{assumption_perturbation}. For example, in the simulations in Section \ref{section_simulation_evidence}, the perturbation $\xi$ is drawn uniformly from $\Xi:=[0, \bar{\varepsilon}]^{d_{\theta}+k}$ with $\bar{\varepsilon}=10^{-3}$. Importantly, the results ahead are not valid for $\bar{\varepsilon}=0$ or $\bar{\varepsilon}\downarrow 0$ as $n\to \infty$. The results also do not hold if $P_{\xi}$ is not absolutely continuous with respect to the Lebesgue measure; this rules out, for instance, non-stochastic perturbations. 

Recall the linear programs \eqref{eq_P_inf_xi_minus} and \eqref{eq_P_inf_xi_plus} from Section \ref{section_main_ideas}. These programs represent ``perturbed'' versions of the linear program \eqref{P_inf}. Since $\varepsilon \in \mathbb{R}_{+}^{k}$, the perturbations to the moment inequalities enlarge the feasible region relative to the linear program \eqref{P_inf}. From the inequality in \eqref{eq_perturbed_inequality}, the value functions of the perturbed linear programs can then be used to construct outer bounds for the value functions for the unperturbed programs.

Let us define a perturbed version of the identified set:
\begin{align}
\Theta_{I}(P,\xi):= \{\theta \in \mathbb{R}^{d_{\theta}} : \E_{P}[m_{j}(W,\theta)]&\leq \varepsilon_{j},\,\, j=1, \ldots, k\}.
\end{align}
Furthermore, let $I(\theta,P,\xi)$ denote the set indexing the binding moment inequalities at $\theta$ for $P \in \mathcal{P}$ and $\xi \in \mathbb{R}^{d_{\theta}}\times \mathbb{R}_{+}^{k}$:
\begin{align}
    I(\theta,P,\xi):= \{ j \in \{1\ldots,k\} : \E_{P}[m_{j}(W,\theta)] = \varepsilon_{j}  \}.
\end{align}
Now let $\theta_{\ell b,-}^{*}$ and $\theta_{\ell b,+}^{*}$ be any optimal solutions to the problems \eqref{eq_P_inf_xi_minus} and \eqref{eq_P_inf_xi_plus}, and let $\theta_{u b,-}^{*}$ and $\theta_{u b,+}^{*}$ be any optimal solutions to the analogous upper bound problems. Let $G(\theta,P,\xi)$ be the matrix formed by vertically stacking the row vectors $\{\nabla_{\theta} \E_{P}[m_{j}(W,\theta)] \}_{j \in I(\theta,P,\xi)}$. We now present a definition of the ``perturbed versions'' of the regularity conditions from Definition \ref{definition_neq_cq_us}.

\begin{figure}[!t]
\centering
\begin{subfigure}[!t]{0.42\textwidth}
\centering
\includegraphics[width=\textwidth]{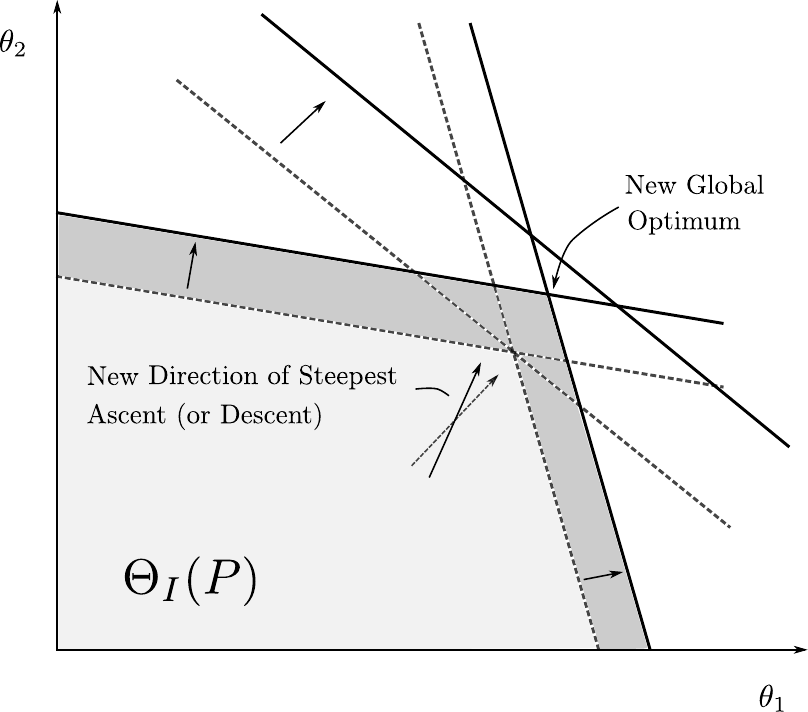}
\caption{This figure illustrates how perturbing the moment conditions can correct a failure of condition (CQ). In particular, relative to Figure \ref{fig_cq_failure}, introducing small perturbations to the problem ensures ($P_{\xi}-$a.s.) that at most two constraints bind. } \label{fig_cq_success}
\end{subfigure}
\hfill
\begin{subfigure}[!t]{0.42\textwidth}
\centering
\includegraphics[width=\textwidth]{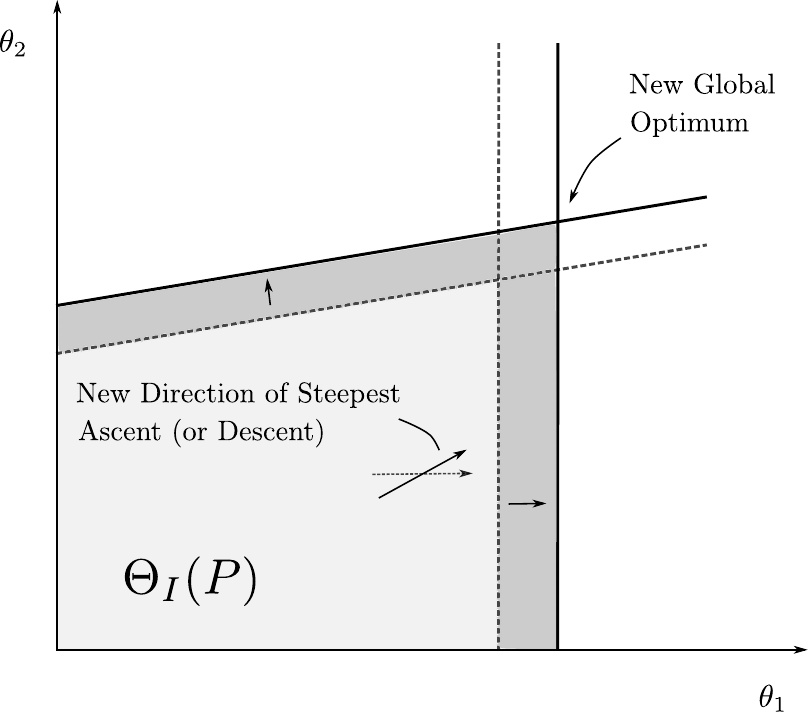}
\caption{This figure illustrates how perturbing the moment conditions and objective function corrects the failure of condition (US). In particular, relative to Figure \ref{fig_us_failure}, perturbing the problem ensures that ($P_{\xi}-$a.s.) the global optimum is always obtained in a ``corner'' of the feasible region.} \label{fig_us_success}
\end{subfigure}
\caption{An illustration of how introducing random perturbations can correct failures of condition (CQ) and (US). This figure should be compared to Figure \ref{fig_failure_cq_us}.}\label{fig_success_cq_us}
\end{figure}

\begin{definition}[Conditions (NE), (CQ) and (US) for the Perturbed Programs]\label{definition_neq_cq_us_perturbed}
Consider the optimization problems \eqref{eq_P_inf_xi_minus} and \eqref{eq_P_inf_xi_plus}, as well as the analogous upper bound problems. We say that the pair $(P,\xi)$ satisfies condition (NE) if $\text{int}(\Theta_{I}(P,\xi))\neq \emptyset$. We say that the pair $(P,\xi)$ satisfies condition (CQ) if:
\begin{align}
\min \left\{ eig\left(G_{\ell b, -}(P,\xi)\right), eig\left(G_{\ell b, +}(P,\xi)\right),eig\left(G_{u b, -}(P,\xi)\right),eig\left(G_{u b,+}(P,\xi)\right) \right\} >0,
\end{align}
where:
\begin{align*}
G_{\ell b, -}(P,\xi) &:=G(\theta_{\ell b,-}^{*},P,\xi)G(\theta_{\ell b,-}^{*},P,\xi)^\top,&& G_{\ell b, +}(P,\xi) :=G(\theta_{\ell b,+}^{*},P,\xi)G(\theta_{\ell b,+}^{*},P,\xi)^\top,\\
G_{u b, -}(P,\xi) &:=G(\theta_{u b,-}^{*},P,\xi)G(\theta_{u b,-}^{*},P,\xi)^\top, &&G_{u b, +}(P,\xi) :=G(\theta_{u b,+}^{*},P,\xi)G(\theta_{u b,+}^{*},P,\xi)^\top,
\end{align*}
and where $eig(A)$ denotes the minimum eigenvalue of the matrix $A$. Finally, we say that the pair $(P,\xi)$ satisfies condition (US) if the optimal solutions $\theta_{\ell b,-}^{*}(P,\xi)$, $\theta_{\ell b,+}^{*}(P,\xi)$, $\theta_{u b,-}^{*}(P,\xi)$ and $\theta_{u b,+}^{*}(P,\xi)$ exist and are unique.
\end{definition}
Definition \ref{definition_neq_cq_us_perturbed} is the analog of Definition \ref{definition_neq_cq_us}, except now all quantities depend on the random perturbation $\xi$. The following result is instrumental to our approach. 
\begin{lemma}\label{lemma_spingarn_rockafeller}
Suppose that Assumptions \ref{assump_continuity} and \ref{assump_regularity} hold. Fix any $P \in \mathcal{P}$ and let $\xi\sim P_{\xi}$ be any random vector satisfying Assumption \ref{assumption_perturbation}. Then the pair $(P,\xi)$ satisfies conditions (NE), (CQ) and (US) $P_{\xi}-$a.s.  
\end{lemma}
The proof of this Lemma uses a result from \cite{spingarn1979generic}, which is in turn derived using Sard's Theorem from differential topology. Lemma \ref{lemma_spingarn_rockafeller} implies that, for any $P \in \mathcal{P}$ satisfying Assumptions \ref{assump_continuity} and \ref{assump_regularity}, conditions (NE), (CQ) and (US) can be made to hold a.s. \kern-0.25em by randomly perturbing the optimization problems in \eqref{P_inf} and \eqref{P_sup}. Figure \ref{fig_success_cq_us} illustrates how the failure of conditions (CQ) and (US) from Figure \ref{fig_failure_cq_us} can be restored after perturbing the objective function and constraints. We also work through two specific examples in Appendix C, showing how uniform Hadamard differentiability can be restored by perturbations. 
Using Lemma \ref{lemma_spingarn_rockafeller}, we obtain a useful corollary to Theorem \ref{thm_uhd}. 
\begin{corollary}\label{corollary_thm_uhd}
Suppose Assumptions \ref{assump_continuity} and \ref{assump_regularity} hold. Fix any $P \in \mathcal{P}$, and let $\xi\sim P_{\xi}$ be any random vector satisfying Assumption \ref{assumption_perturbation}. Consider the value functions $\Psi_{I,-}^{\ell b}(\,\cdot\,,\xi)$ and $\Psi_{I,+}^{\ell b}(\,\cdot\,,\xi)$ defined by the programs \eqref{eq_P_inf_xi_minus} and \eqref{eq_P_inf_xi_plus}. Then there exists a set $E\subset\Xi$, possibly depending on $P$, with $P_{\xi}(E)=1$ such that for all $\xi \in E$, the value functions $\Psi_{I,-}^{\ell b}(\,\cdot\,,\xi)$ and $\Psi_{I,+}^{\ell b}(\,\cdot\,,\xi)$ are uniformly Hadamard differentiable at $P$ tangential to $\mathcal{T}_{P}(\mathcal{F})$. 
\end{corollary}
Corollary \ref{corollary_thm_uhd} gives us a way to ensure the value function of a linear program is UHD $P_{\xi}-$a.s., and is used to construct confidence sets in the next section. Although Corollary \ref{corollary_thm_uhd} is stated for $\Psi_{I,-}^{\ell b}(\,\cdot\,,\xi)$ and $\Psi_{I,+}^{\ell b}(\,\cdot\,,\xi)$, the same result holds for the upper bound perturbed value functions. Again, since the result might fail to hold for $\xi$ in some $P_{\xi}-$null set $\Xi_{0} \subset \Xi$, stochastic perturbations are essential.

\subsection{From Differentiability to Confidence Sets}\label{subsection_confidence_sets}

We now construct a confidence set for the true but partially identified value of a functional or scalar subvector of interest using the differentiability results of the previous sections. Our proposed procedure uses the nonparametric bootstrap, which we formalize as an assumption.  

\begin{assumption}\label{assump_bootstrap}
The bootstrap samples $\{W_{i}^{b}\}_{i=1}^{n}$, for $b=1,\ldots, \overline{B}$, are drawn i.i.d. with replacement with equal probability from the original sample $\{W_{i}\}_{i=1}^{n}$. 
\end{assumption} 
Our next result uses a functional delta method to approximate the distribution of the value functions of our perturbed linear programs. We state the result only for the perturbed program \eqref{eq_P_inf_xi_minus}, although it also holds without substantial modification for the perturbed program \eqref{eq_P_inf_xi_plus}, as well as for the analogous upper bound perturbed programs.

\begin{lemma}\label{lemma_stopro_uniform1_intext}
Suppose Assumptions \ref{assump_continuity}, \ref{assump_regularity} and \ref{assumption_perturbation} hold and consider program \eqref{eq_P_inf_xi_minus}. Fix any sequence $\{P_{n}\} \subset \mathcal{P}$ such that $P_{n}\rsq P$ in $\mathcal{P}$, and let $\G_{n,P_{n}} := \sqrt{n}(\P_{n} - P_{n})$. Suppose that $\mathbb{G}_{n,P_{n}} \rsq \mathbb{G}_{P}$ in $\ell^{\infty}(\mathcal{F})$ over the sequence $P_{n} \rsq P$ in $\mathcal{P}$, where $\mathbb{G}_{P}$ is a tight Borel measurable element in $\mathcal{T}_{P}(\mathcal{F})$. Then:
\begin{align}
\sqrt{n}(\Psi_{I,-}^{\ell b}(\mathbb{P}_{n},\xi) - \Psi_{I,-}^{\ell b}(P_{n},\xi))  \rsq (\Psi_{I,-}^{\ell b})_{P,\xi}'(\mathbb{G}_{P}),\label{lower_stopro_uniform_intext}
\end{align}
$P_{\xi}-$a.s., where $(\Psi_{I,-}^{\ell b})_{P,\xi}'(\,\cdot\,)$ is the uniform Hadamard derivative from Corollary \ref{corollary_thm_uhd}. Furthermore, let $\mathbb{G}_{n}^{b}:= \sqrt{n}(\mathbb{P}_{n}^{b} - \mathbb{P}_{n})$ and suppose that $\mathbb{G}_{n}^{b}\overset{p}{\underset{M}{\rsq}} \mathbb{G}_{P}$ over the sequence $P_{n} \rsq P$ in $\mathcal{P}$, where ``$\overset{p}{\underset{M}{\rsq}}$'' denotes conditional (on $\{W_{i}\}_{i=1}^{n}$) weak convergence in probability in $\ell^\infty(\mathcal{F})$.\footnote{See Definition A.4 in Appendix A.} Then under Assumptions \ref{assump_continuity}, \ref{assump_regularity}, \ref{assumption_perturbation} and \ref{assump_bootstrap}:
\begin{align}
\sqrt{n}(\Psi_{I,-}^{\ell b}(\mathbb{P}_{n}^{b},\xi) - \Psi_{I,-}^{\ell b}(\P_{n},\xi)) \overset{p}{\underset{M}{\rsq}} (\Psi_{I,-}^{\ell b})_{P,\xi}'(\G_{P}),\label{lower_stopro_uniform_bootstrap_intext}
\end{align}
$P_{\xi}-$a.s.  
\end{lemma} 

Lemma \ref{lemma_stopro_uniform1_intext} shows that the nonparametric bootstrap can be used to approximate the limiting distributions of the value functions of our perturbed linear programs, and immediately suggests a method of inference. Recall the ``maximally relaxed'' version of the identified set $\overline{\Psi}_{I}(P)$ from \eqref{eq_maximally_relaxed}. The following is our main theoretical result.
\begin{theorem}\label{corollary_uniformity}
Suppose Assumptions \ref{assump_continuity}, \ref{assump_regularity}, \ref{assumption_perturbation} and \ref{assump_bootstrap} hold. Then the confidence set $C_{n}^{\psi}(1-\alpha,\xi)$ in \eqref{CS_proposal} satisfies:
\begin{align}
\liminf_{n \to \infty} \inf_{\{ (\psi,P): \:\: \psi \in \Psi_{I}(P), \:\:P \in \mathcal{P} \}} (\text{Pr}_{P}\times P_{\xi})(\psi \in C_{n}^{\psi}(1-\alpha,\xi)) = 1.\label{eq_main_theorem_1}
\end{align}
Furthermore, for any $0<\alpha<1$, there exists a set $\widetilde{\Psi}_{\alpha}(P)$ with $\Psi_{I}(P) \subset \widetilde{\Psi}_{\alpha}(P) \subset \overline{\Psi}_{I}(P)$ such that:
\begin{align}
\liminf_{n \to \infty} \inf_{\{ (\psi,P): \:\: \psi \in \widetilde{\Psi}_{\alpha}(P), \:\:P \in \mathcal{P} \}} (\text{Pr}_{P}\times P_{\xi})(\psi \in C_{n}^{\psi}(1-\alpha,\xi)) \geq 1-\alpha,\label{eq_main_theorem_2}
\end{align}
with equality holding in \eqref{eq_main_theorem_2} if either $\Psi_{I}^{\ell b}(P,\xi)$ or $\Psi_{I}^{u b}(P,\xi)$ are continuous at their ($P_{\xi}$) $\alpha-$quantiles for some $P\in \mathcal{P}$. 
\end{theorem}
The result shows that, after introducing certain random perturbations, a simple nonparametric bootstrap procedure can be used to construct a uniformly valid confidence set for a partially identified functional. The first part of Theorem \ref{corollary_uniformity} shows that the proposed confidence set is uniformly asymptotically valid, but very conservative. Intuitively, this comes from the use of a global and non-vanishing perturbation, which implies that our confidence set will never collapse to the identified set, but instead will collapse to an outer set. Any tests based on inverting our confidence set will also have zero power for local alternatives drifting to a point in the identified set. However, the power will generally be non-zero for points outside the identified set.\footnote{
	In particular, this is also true for points outside $\Psi_{I}(P)$ but inside $\Psi_{I}(P,\xi)$ for a fixed perturbation draw $\xi$. To see the intuition, note that a fixed alternative $\psi^\dagger$ may lie outside $\Psi_{I}(P)$ and $\Psi_{I}(P,\xi_{1})$ but inside $\Psi_{I}(P,\xi_{2})$ for two different draws $\xi_{1}$ and $\xi_{2}$ from $P_{\xi}$. Studying the coverage of $\psi^\dagger$ by our confidence set requires taking repeated draws from $P_{\xi}$; intuitively, if $\psi^\dagger$ is outside $\Psi_{I}(P,\xi)$ for sufficiently many draws $\xi$ from $P_{\xi}$, our procedure will not have zero power at $\psi^\dagger$.
} 
The second part of Theorem \ref{corollary_uniformity} shows that the confidence set obtains exact uniform asymptotic coverage over a set $\widetilde{\Psi}_{\alpha}(P)$ that lies between the identified set $\Psi_{I}(P)$ and the maximally perturbed identified set $\overline{\Psi}_{I}(P)$. When $\overline{\Psi}_{I}(P)$ is close to $\Psi_{I}(P)$, this implies that the confidence set obtains exact coverage over a small expansion of the identified set. In this respect, the approach modifies the coverage objective slightly in exchange for computational tractability. While the choice of $\alpha$ does not influence the coverage result in \eqref{eq_main_theorem_1}, it will influence the set $\widetilde{\Psi}_{\alpha}(P)$ satisfying \eqref{eq_main_theorem_2}. The proof of the result proceeds by explicitly constructing a set $\widetilde{\Psi}_{\alpha}(P)$ satisfying the conditions stated in the Theorem. The proof shows that for any $\alpha_{1}<\alpha_{2}$ we have $\widetilde{\Psi}_{\alpha_{1}}(P) \subset\widetilde{\Psi}_{\alpha_{2}}(P)$; in other words, for an increasing sequence of $\alpha$'s, the outer sets that are covered exactly by our confidence set are also increasing. 

In the end, the approach moves the goalpost, trading off exact coverage on an outer set for a confidence set that is valid under weak assumptions, and is computationally easy to construct. The description of our final recommended procedure can be found in Section \ref{section_main_ideas}.

\section{Practical Considerations and Simulation Evidence}\label{section_simulation_evidence}

\subsection{On the Magnitude of the Perturbation}\label{section_ebar_choice}

The main tuning parameter in our method is $\bar{\varepsilon}$, which governs the upper bound on the support of the perturbations for both the objective function and constraints. Theorem \ref{corollary_uniformity} shows that, at least asymptotically, any choice of $\bar{\varepsilon}>0$ is sufficient to guarantee the coverage results in \eqref{eq_main_theorem_1} and \eqref{eq_main_theorem_2}. This seems to suggests that $\bar{\varepsilon}>0$ should be chosen as small as possible, since this will produce the shortest confidence interval using our procedure. In that case, \eqref{eq_main_theorem_2} in Theorem \ref{corollary_uniformity} also implies our confidence set has exact coverage on an outer set $\widetilde{\Psi}_{\alpha}(P)$ that will be very close to the identified set $\Psi_{I}(P)$. However, this reasoning can be misleading, since the choice of $\bar{\varepsilon}>0$ can have important impacts on the coverage of our confidence set in finite samples. In this subsection we explore the tradeoff between power and coverage associated with various choices of $\bar{\varepsilon}$ in finite samples. We then discuss the impact of taking multiple draws from the perturbation distribution, a practice we discourage. Throughout this section, $\xi$ is drawn from the uniform distribution on $[0,\bar{\varepsilon}]^{d_{\theta}+k}$ and $b_{n}=1/\sqrt{\log(n)}$.\footnote{The length of the identified set was large enough so that the quantile correction in \eqref{eq_gamma_correction} in Step 5 of the procedure did not play a substantial role in this example.}  

To gain insight into the role of $\bar{\varepsilon}$ in finite samples, we consider a simple generalization of a DGP analyzed by \cite{andrews2009invalidity}.\footnote{This example is also considered in \cite{gafarov2021inference} Section 3.3.2. } In particular, let:
\begin{align*}
(W_{i1}\,\,
W_{i2}\,\,
\ldots\,\,
W_{i2L})^\top\sim 
N \left( 0,I_{L} \otimes  \Sigma \right), \quad \text{ where }\quad  \Sigma :=
\begin{pmatrix} 
1 & -0.99\\
-0.99 & 1
\end{pmatrix}.
\end{align*}
Now set $\psi(W_{i},\theta) = \psi(\theta) = \theta$ (a scalar) and consider the following optimization problems:
\begin{align}
\Psi_{I}^{\ell b}(P)/\Psi_{I}^{ub}(P) = \underset{\theta \in [-1,1]}{\min/\max} \,\,\theta, \quad\text{ s.t. }\quad \theta \geq  \E_{P}[W_{i\ell}], \,\,\ell=1,\ldots,2L.\label{eq_CQ_failure}
\end{align}
It is straightforward to see that:
\begin{align}
\Psi_{I}(P)=\left[\max_{\ell=1,\ldots, 2L}\E_{P}[W_{i\ell}],\,\,1\right].\label{eq_simulation_identified_set}
 \end{align}
In particular, $P \mapsto \Psi_{I}^{\ell b}(P)$ is not Hadamard differentiable (and thus not UHD) under this DGP, since condition (CQ) is violated in the minimization problem. In finite samples, we expect the distortions in coverage when $\bar{\varepsilon}=0$ to be the largest when $L$ is large. We consider the case when $L=1$---which matches the DGP in \cite{andrews2009invalidity}---and $L=10$. The latter DGP will emphasize the severe under-coverage that is possible in this simple example. 

\begin{figure}[!t]
\centering
\begin{subfigure}[!t]{0.49\textwidth}
\centering
\includegraphics[width=\textwidth]{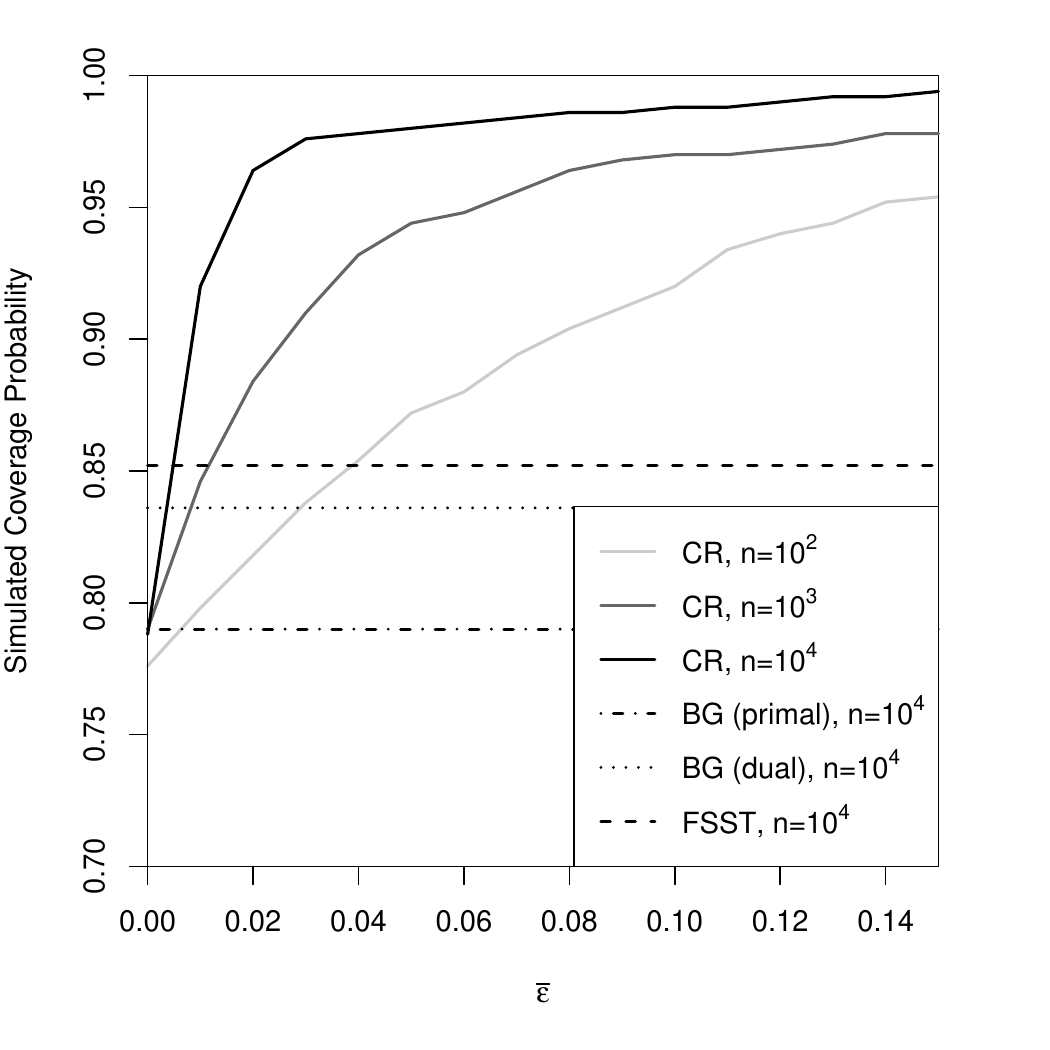}
\caption{$L=1$.}
\end{subfigure}
\hfill
\begin{subfigure}[!t]{0.49\textwidth}
\centering
\includegraphics[width=\textwidth]{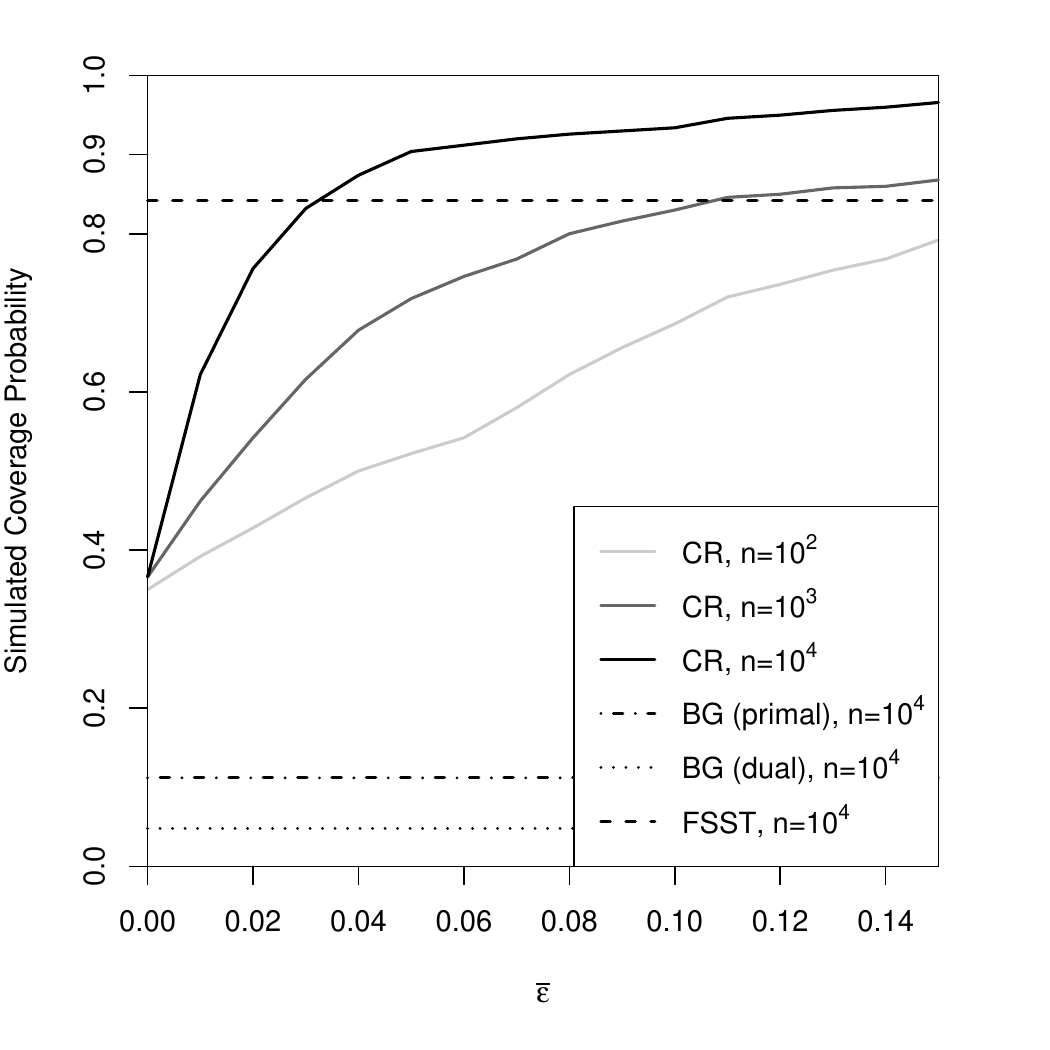}
\caption{$L=10$.}
\end{subfigure}
\caption{The $90\%$ coverage probability for our confidence set (labelled CR) across $R=500$ replications for the example in \eqref{eq_CQ_failure} with $L\in \{1,10\}$ for various $\bar{\varepsilon}$ and various sample sizes. Also included are coverage probabilities for the \cite{gafarov2021inference} confidence set (labelled BG), and the \cite{fang2021inference} confidence set (FSST), both at $n=10^4$. All results are with $\overline{B}=999$ bootstrap samples.}\label{fig_cq_epsbar}
\end{figure}

Figure \ref{fig_cq_epsbar}(a) shows the coverage probability of our confidence set as a function of $\bar{\varepsilon}$ for $\alpha=0.1$, for various samples sizes, in the case when $L=1$. Figure \ref{fig_cq_epsbar}(b) shows the case when $L=10$. For comparison, we also include the coverage probability at $n=10^{4}$ using the \cite{fang2021inference} procedure, and both the primal and dual formulations of the \cite{gafarov2021inference} procedure. The dual formulation is explictly recommended by \cite{gafarov2021inference} for this DGP, since the constraint qualification assumptions in \cite{gafarov2021inference} fail for the primal formulation but hold in the dual formulation.\footnote{C.f. \cite{gafarov2021inference} section 3.3.2. } Since it is not always easy to tell when these assumptions fail, the primal formulation is included for comparison.  Consistent with the discussion above, when $\bar{\varepsilon}=0$ the $90\%$ coverage probability is $0.79$ when $L=1$ and $0.408$ when $L=10$. The rate at which the coverage probability improves as $\bar{\varepsilon}$ becomes larger depends on the sample size. This is consistent with the theoretical predictions, which suggest that even small values of $\bar{\varepsilon}$ will lead to over-coverage in sufficiently large samples. However, it also illustrates a tradeoff in finite samples: as emphasized in Figure \ref{fig_cq_epsbar}(b), when $\bar{\varepsilon}$ is ``small'' relative to sampling uncertainty, the perturbation may not be sufficient to correct the distortions in the coverage probability when the (unperturbed) value functions are not UHD. On the other hand, ``large'' values of $\bar{\varepsilon}$ relative to sampling uncertainty can lead to over-coverage in finite samples. Note that Figure \ref{fig_cq_epsbar} is also informative about the effect of rescaling all moment inequalities by a positive constant, which is essentially equivalent to a rescaling of the value of $\bar{\varepsilon}$.\footnote{When each moment condition is re-scaled by a different amount, the effect in finite sample can be ambiguous. In unreported simulations, we found that re-scaling the variance of only $W_{i1}$ (for $L=1$ and $L=10$) had essentially no effect on the coverage probability. }

In comparison, the confidence sets of both \cite{fang2021inference} and \cite{gafarov2021inference} under-cover slightly for $n=10^4$ when $L=1$: the $90\%$ coverage probability for the \cite{fang2021inference} procedure was $0.852$, and the $90\%$ coverage probability for the \cite{gafarov2021inference} procedure was $0.790$ for the primal and $0.836$ for the dual. The \cite{fang2021inference} procedure has similar coverage when $L=10$, at $0.842$. For the \cite{gafarov2021inference} procedure, the $90\%$ coverage probability when $L=10$ is only $0.112$ for the primal formulation and $0.048$ for the dual formulation. Although the assumptions of \cite{gafarov2021inference} hold for the dual formulation, the low coverage appears to result from the failure of the \cite{gafarov2021inference} bias correction procedure when using the recommended tuning parameters.\footnote{All of the \cite{gafarov2021inference} results were obtained using the tuning parameters $\mu_{n}$ and $\kappa_{n}$ recommended in Section 5.2 of \cite{gafarov2021inference}. The coverage probability using the \cite{gafarov2021inference} procedure improved only when the values of the tuning parameters were substantially increased. However, as illustrated in this section, our procedure also faces the same difficult problem of selecting appropriate tuning parameters for good finite sample performance. }  For $n=10^4$, our procedure obtains nominal coverage at $\bar{\varepsilon}=0.008$ when $L=1$ and $\bar{\varepsilon}=0.049$ when $L=10$.

\begin{figure}[!t]
\centering
\begin{subfigure}[!t]{0.49\textwidth}
\centering
\includegraphics[width=\textwidth]{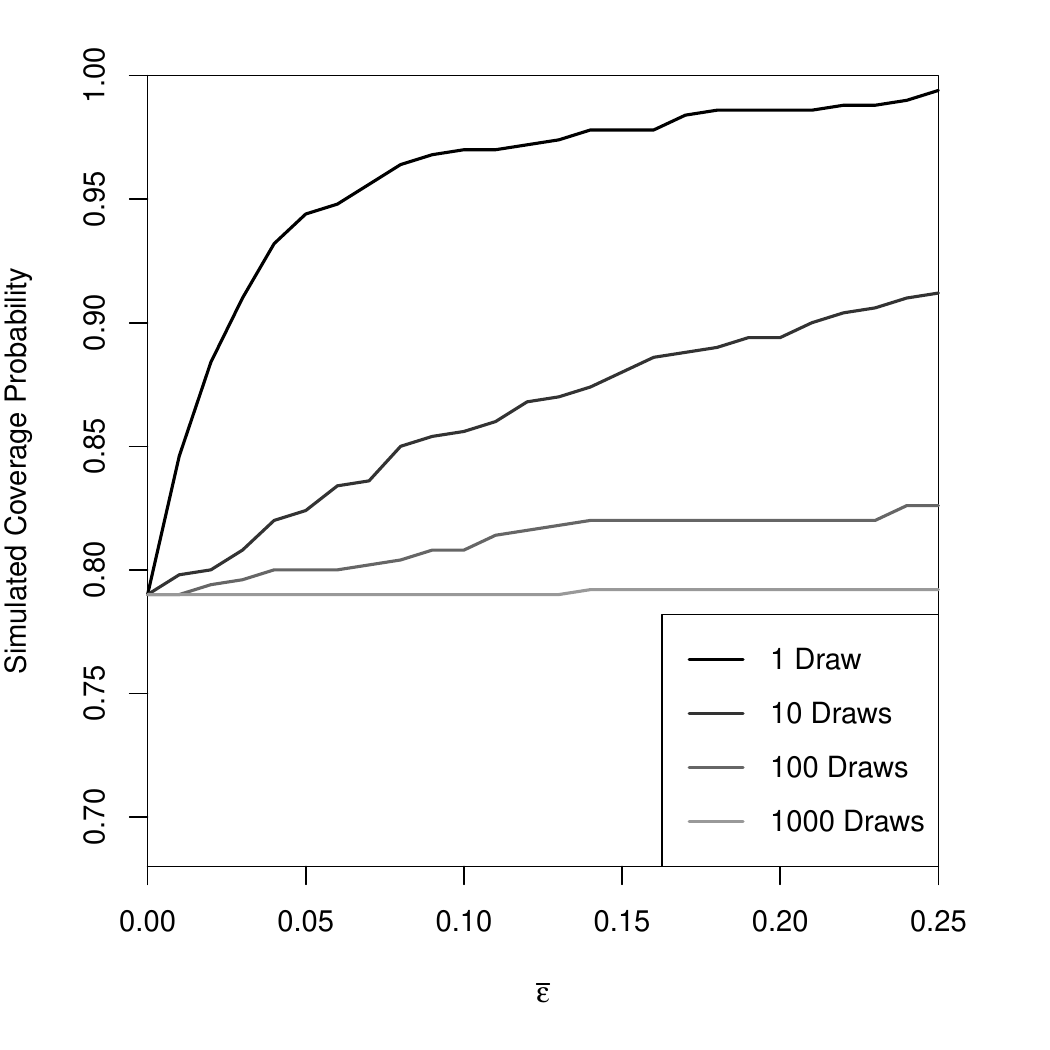}
\caption{$L=1$.}
\end{subfigure}
\hfill
\begin{subfigure}[!t]{0.49\textwidth}
\centering
\includegraphics[width=\textwidth]{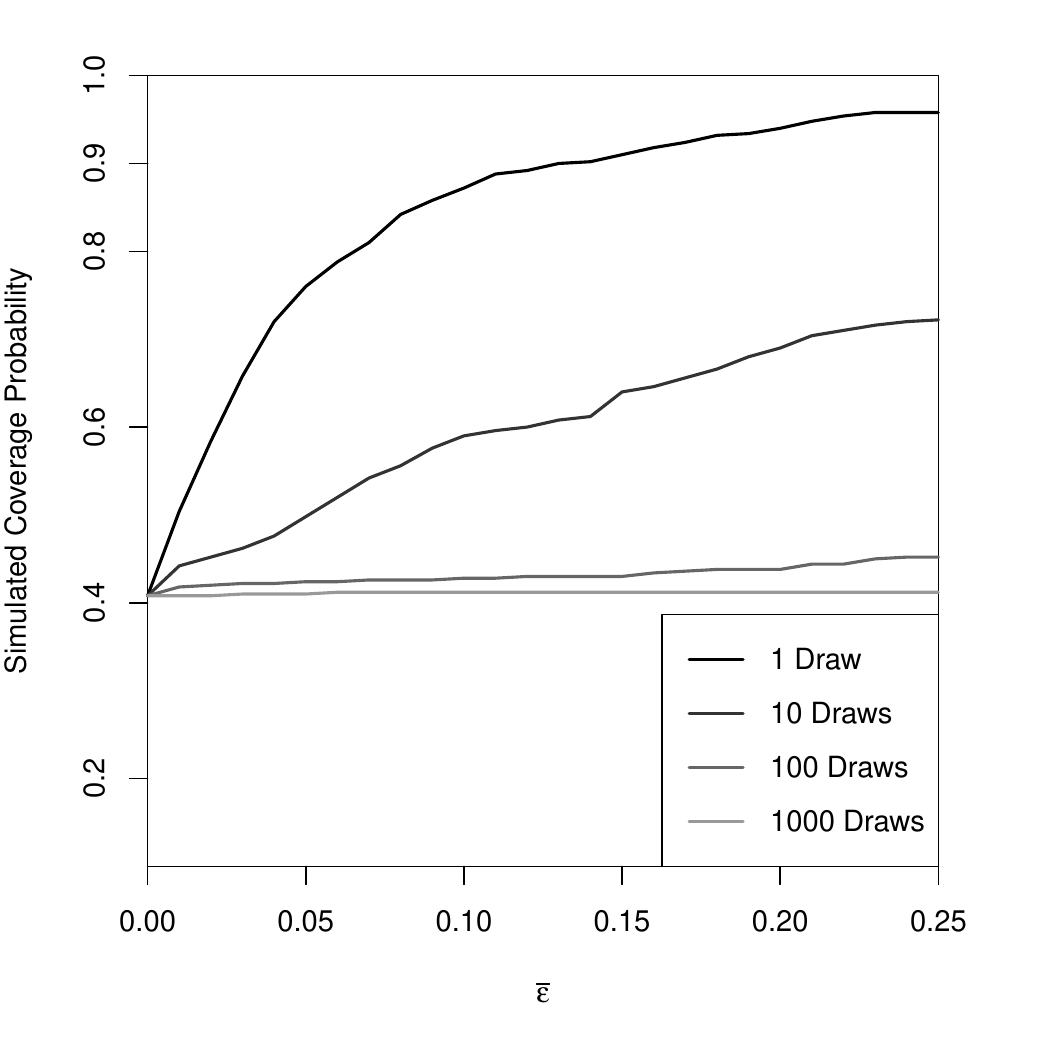}
\caption{$L=10$. }
\end{subfigure}
\caption{The $90\%$ coverage probability for our confidence set across $R=500$ replications for the example in \eqref{eq_CQ_failure} with $n=10^3$ and $L\in\{1,10\}$. The coverage probability is computed by taking multiple draws of the perturbation, and selecting the confidence set with the smallest length across all draws. All results are with $\overline{B}=999$ bootstrap samples. }\label{fig_gaming}
\end{figure}

In addition to exploring the support of the perturbation, we also explore the consequences of drawing multiple perturbations with a simulation exercise. Our ultimate goal with this exercise is to discourage the practice of taking multiple perturbation draws. Since the length of our confidence set depends on the random draw $\xi$, for a fixed $\bar{\varepsilon}$ researchers may be tempted to draw multiple perturbations until a confidence set of favourable length is obtained.  With larger values of $\bar{\varepsilon}$, redrawing the perturbation can have a large effect on the length of the confidence set. Although redrawing the perturbation a finite number of times is asymptotically valid for our procedure, it can have important impacts in finite sample. 

To explore the impact of taking multiple perturbation draws, we again consider the example in \eqref{eq_CQ_failure}. For both $L=1$ and $L=10$, we consider redrawing the perturbation for the moment inequalities $1$, $10$, $100$ and $1000$ times, and take the smallest confidence set across all draws. Figure \ref{fig_gaming} shows the $90\%$ coverage probability as a function of $\bar{\varepsilon}$ obtained from the various confidence sets using this procedure. While the coverage probability is increasing in $\bar{\varepsilon}$ for all confidence sets, the figure shows that---for a fixed $\bar{\varepsilon}$---taking many perturbation draws and selecting the most favorable draw may limit the ability of the perturbation to overcome coverage distortions due to UHD failures in finite sample. With $1000$ perturbation draws, the benefits of the perturbation are eliminated almost entirely. In the end, we recommend that researchers use the uniform distribution and take a single draw for the perturbation.

\subsection{Additional Simulation Evidence}

To provide additional evidence, we perform Monte Carlo experiments on the three empirical examples introduced in Section \ref{section_main_ideas}. The examples show that the coverage is at least the nominal level for all samples sizes, and illustrate the conservative distortion that is possible in large samples. Throughout this section we also compare with the approaches of \cite{fang2021inference} and \cite{gafarov2021inference}. The details on all the DGPs considered in the simulations are provided in Appendix D.  In all Monte Carlo exercises we take $\overline{B}=999$ bootstrap samples for each experiment. We implement each experiment $500$ times to compute the simulated coverage probability, which is determined by computing the worst case coverage probability across all points in the identified set. We consider various sample sizes $n\in \{250,500,10^3,10^4,10^5,10^6\}$. The large sample sizes are considered to illustrate the conservative nature of our confidence set, but also to alleviate concerns that our approach delivers confidence sets that are overly-wide relative to the identified set in large samples. For each DGP, $\xi$ is drawn from the uniform distribution on $[0,\bar{\varepsilon}]^{d_{\theta}+k}$ with $\bar{\varepsilon} = 10^{-3}$, and $b_{n}=1/\sqrt{\log(n)}$. In Appendix D.4 we explore the sensitivity of our results to the choice of $\bar{\varepsilon}$, and also explore the role of $\gamma(\alpha)$ from \eqref{eq_gamma_correction}.

\begin{sidewaystable}
\begin{threeparttable}
  \centering
  \caption{This table shows the results of the simulation exercise for the missing data example with $\overline{B}=999$ bootstrap replications for each experiment, where $500$ experiments were run to determine the coverage probability. The parameter of interest is the unconditional average of $Y$ where $Y \in \{1,2,3,4,5\}$ and where $Y$ is observed only when $D=1$. The true identified set is $[2.8,3.2]$.}
  \begin{footnotesize}
    \begin{tabular}{|l|cc|cc|ccc|ccc|ccc|}
    \toprule
    \toprule
                &\multicolumn{2}{c}{Avg. Estimate} &  \multicolumn{2}{|c|}{Avg. Perturbed} & \multicolumn{3}{c|}{$1-\alpha = 0.90$} & \multicolumn{3}{c|}{$1-\alpha =0.95$} & \multicolumn{3}{c|}{$1-\alpha =0.99$} \\
    \midrule
    \multicolumn{1}{|c|}{Sample Size} & LB & UB & LB & UB & Coverage & Avg. LB & Avg. UB & Coverage & Avg. LB & Avg. UB & Coverage & Avg. LB & Avg. UB  \\
     \midrule
    $n=250$ & 2.808&	3.201&	2.795&	3.216&	0.932&	2.659&	3.352&	0.976&	2.628&	3.384&	0.994&	2.570&	3.442\\
    $n=500$ & 2.797&	3.199&	2.783&	3.214&	0.962&	2.692&	3.305&	0.980&	2.669&	3.329&	0.998&	2.627&	3.371\\
    $n=10^3$& 2.801&	3.201&	2.787&	3.217&	0.954&	2.726&	3.278&	0.978&	2.710&	3.294&	1.000&	2.679&	3.325\\
    $n=10^4$& 2.801&	3.200&	2.787&	3.216&	0.970&	2.768&	3.235&	0.990&	2.763&	3.240&	0.998&	2.753&	3.250\\
    $n=10^5$& 2.800&	3.200&	2.787&	3.215&	1.000&	2.781&	3.221&	1.000&	2.779&	3.223&	1.000&	2.776&	3.226\\
    $n=10^6$& 2.800&  	3.200&  2.787&  3.216& 	1.000& 	2.785& 	3.217& 	1.000& 	2.784& 	3.218& 	1.000& 2.783& 3.219\\
    \bottomrule
    \bottomrule
    \end{tabular}%
    \end{footnotesize}
    \begin{tablenotes}
      \tiny
      \item ``Avg. Estimate'' average value of the lower and upper bounds of the identified set. ``Avg. Perturbed'' is the corresponding average lower and upper bounds for the set $\Psi_{I}(P,\xi)$. ``Coverage'' refers to the worst case coverage probability across all points in the identified set.  
    \end{tablenotes} 
  \label{table_missing_data}%
  \end{threeparttable}~\\~\\

\begin{threeparttable}
  \centering
  \caption{This table shows the results of the simulation exercise for the interval valued regression example with $\overline{B}=999$ bootstrap replications for each experiment, where $500$ experiments were run to determine the coverage probability. The parameter of interest is the first component, $\theta_{1}$, of the vector $\theta$, where $Y=X^\top\theta + \varepsilon$, and where $Y$ is interval-valued with $P(Y_{*}\leq Y \leq Y^{*})=1$. The true identified set is $[1.11,1.19]$.}
  \begin{footnotesize}
    \begin{tabular}{|l|cc|cc|ccc|ccc|ccc|}
    \toprule
    \toprule
                &\multicolumn{2}{c}{Avg. Estimate} &  \multicolumn{2}{|c|}{Avg. Perturbed} & \multicolumn{3}{c|}{$1-\alpha = 0.90$} & \multicolumn{3}{c|}{$1-\alpha =0.95$} & \multicolumn{3}{c|}{$1-\alpha =0.99$} \\
    \midrule
    \multicolumn{1}{|c|}{Sample Size} & LB & UB & LB & UB & Coverage & Avg. LB & Avg. UB & Coverage & Avg. LB & Avg. UB & Coverage & Avg. LB & Avg. UB  \\
     \midrule
    $n=250$  & 1.100 & 1.180 & 1.099 & 1.182 & 0.940 & 0.912 & 1.368 & 0.968 & 0.877 & 1.404 & 0.994 & 0.808 & 1.471\\
    $n=500$  & 1.114 & 1.194 & 1.113 & 1.195 & 0.938 & 0.982 & 1.327 & 0.974 & 0.957 & 1.351 & 0.992 & 0.909 & 1.399\\
    $n=10^3$ & 1.105 & 1.186 & 1.104 & 1.187 & 0.944 & 1.011 & 1.280 & 0.980 & 0.994 & 1.298 & 0.992 & 0.960 & 1.332\\
    $n=10^4$ & 1.111 & 1.190 & 1.109 & 1.192 & 0.948 & 1.080 & 1.221 & 0.978 & 1.074 & 1.227 & 0.986 & 1.064 & 1.237\\
    $n=10^5$ & 1.110 & 1.190 & 1.109 & 1.192 & 0.964 & 1.100 & 1.201 & 0.978 & 1.098 & 1.203 & 0.994 & 1.094 & 1.206\\
    $n=10^6$ & 1.110 & 1.190 & 1.109 & 1.191 & 0.976 & 1.106 & 1.194 & 0.986 & 1.105 & 1.195 & 0.992 & 1.104 & 1.196\\ 
    \bottomrule
    \bottomrule
        \end{tabular}%
    \end{footnotesize}
    \begin{tablenotes}
      \tiny
      \item ``Avg. Estimate'' average value of the lower and upper bounds of the identified set. ``Avg. Perturbed'' is the corresponding average lower and upper bounds for the set $\Psi_{I}(P,\xi)$. ``Coverage'' refers to the worst case coverage probability across all points in the identified set.   
    \end{tablenotes} 
  \label{table_linear_regression_interval}%
  \end{threeparttable}~\\~\\
\color{black}
\begin{threeparttable}
  \centering
  \caption{This table shows the results of the simulation exercise for the counterfactual policy example in \cite{kasy2016partial} with $\overline{B}=999$ bootstrap replications for each experiment, where $500$ experiments were run to determine the coverage probability. The parameter of interest is $\E_{P}[Y^{A}-Y^{B}]$, which is the difference in the expected treatment effect under two competing policies. The true identified set is $[0.054,0.264]$.}
  \begin{footnotesize}
    \begin{tabular}{|l|cc|cc|ccc|ccc|ccc|}
    \toprule
    \toprule
                &\multicolumn{2}{c}{Avg. Estimate} &  \multicolumn{2}{|c|}{Avg. Perturbed} & \multicolumn{3}{c|}{$1-\alpha = 0.90$} & \multicolumn{3}{c|}{$1-\alpha =0.95$} & \multicolumn{3}{c|}{$1-\alpha =0.99$} \\
    \midrule
    \multicolumn{1}{|c|}{Sample Size} & LB & UB & LB & UB & Coverage & Avg. LB & Avg. UB & Coverage & Avg. LB & Avg. UB & Coverage & Avg. LB & Avg. UB  \\
     \midrule
    $n=250$ & 0.056& 0.264& 0.052 & 0.268 & 0.960& 0.004& 0.306& 0.982&-0.005& 0.314& 0.996 & -0.022& 0.329\\
    $n=500$ & 0.056& 0.265& 0.053 & 0.268 & 0.964& 0.018& 0.296& 0.982& 0.012& 0.301& 0.996 &  0.000& 0.311\\
    $n=10^3$& 0.055& 0.265& 0.052 & 0.268 & 0.970& 0.028& 0.287& 0.986& 0.023& 0.291& 0.998 &  0.014& 0.298\\
    $n=10^4$& 0.055& 0.264& 0.051 & 0.268 & 0.980& 0.044& 0.274& 0.986& 0.042& 0.275& 0.998 &  0.039& 0.277\\
    $n=10^5$& 0.055& 0.264& 0.051 & 0.268 & 1.000& 0.049& 0.270& 1.000& 0.048& 0.270& 1.000 &  0.048& 0.271\\
    $n=10^6$& 0.055& 0.264& 0.051 & 0.268 & 1.000& 0.051& 0.268& 1.000& 0.050& 0.268& 1.000 &  0.050& 0.269\\
    \bottomrule
    \bottomrule
    \end{tabular}%
    \end{footnotesize}
    \begin{tablenotes}
      \tiny
      \item ``Avg. Estimate'' average value of the lower and upper bounds of the identified set. ``Avg. Perturbed'' is the corresponding average lower and upper bounds for the set $\Psi_{I}(P,\xi)$. ``Coverage'' refers to the worst case coverage probability across all points in the identified set.  
    \end{tablenotes} 
  \label{table_counterfactual_policy}%
  \end{threeparttable}
\end{sidewaystable}%

\begin{figure}[!t]
\centering
\begin{subfigure}[!t]{0.49\textwidth}
\centering
\includegraphics[width=\textwidth]{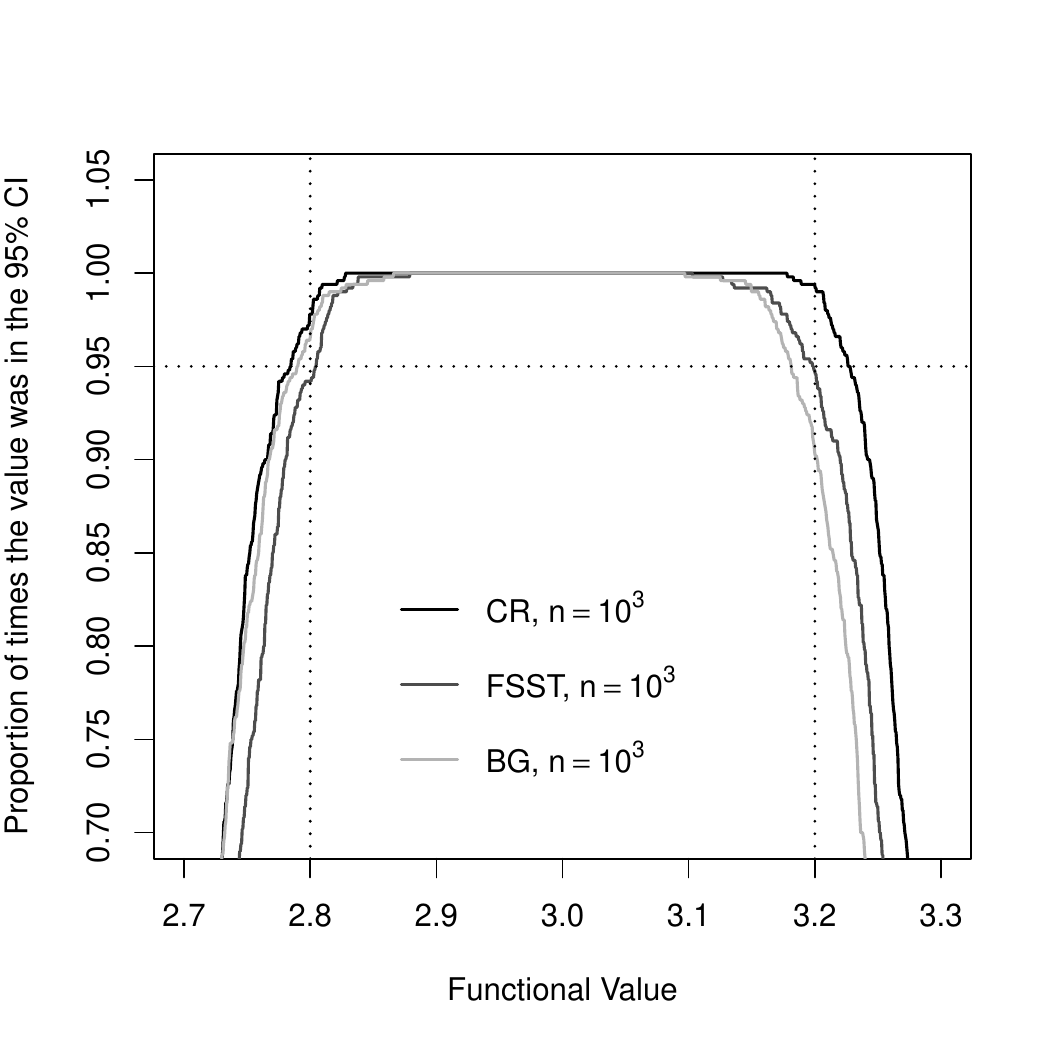}
\caption{} \label{fig_fsst_md_1000}
\end{subfigure}\hfill
\begin{subfigure}[!t]{0.49\textwidth}
\includegraphics[width=\textwidth]{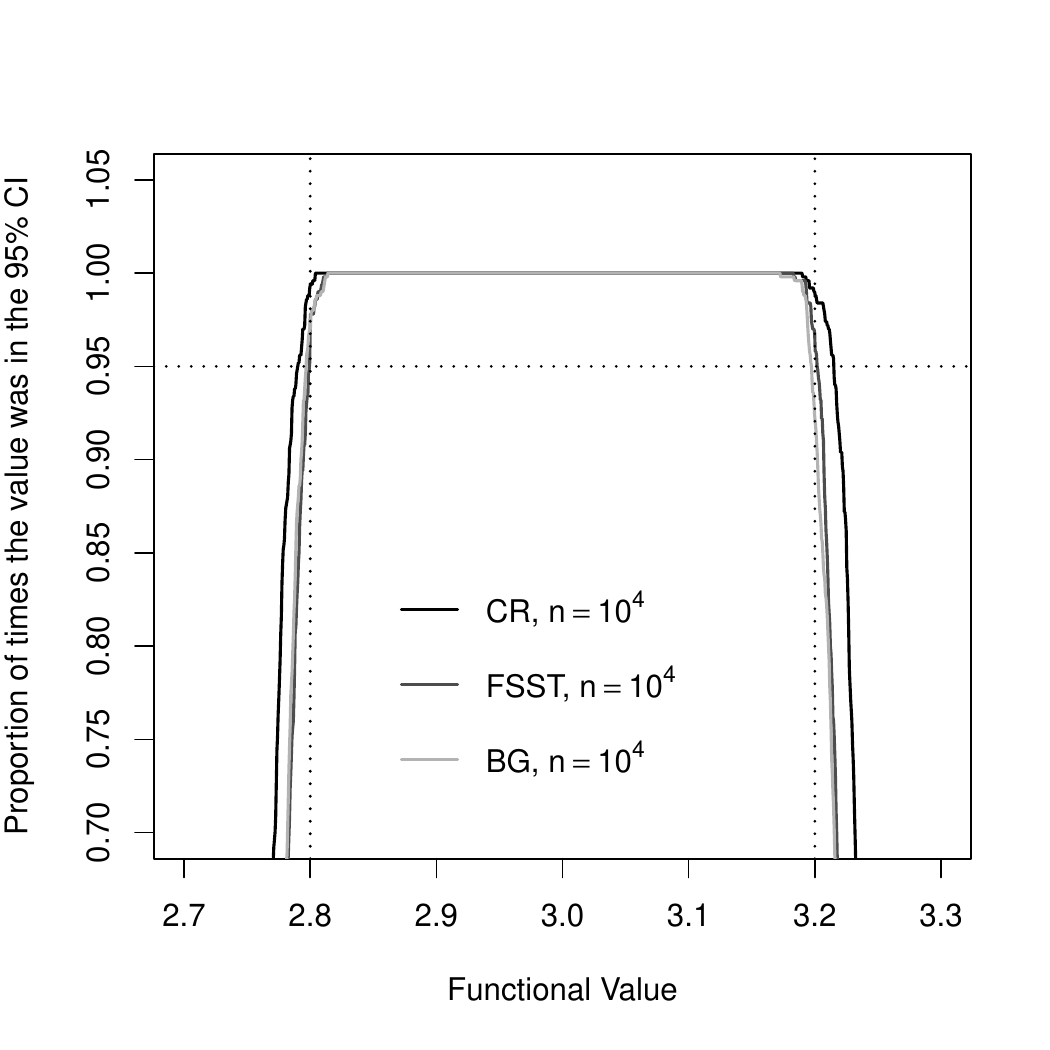}
\caption{} \label{fig_fsst_md_10000}
\end{subfigure}
\caption{
		A comparison between our proposed procedure and the procedure in \cite{fang2021inference} and \cite{gafarov2021inference} for the missing data example. Each subfigure displays the proportion of times various values of the functional $\psi$ associated with the missing data example fell within the $95\%$ confidence set for our proposed procedure (labelled as CR), for the procedure of \cite{fang2021inference} (labelled as FSST), and for the procedure of \cite{gafarov2021inference} (labelled as BG). Figure \ref{fig_fsst_md_1000} shows the case with $n=10^3$, and Figure \ref{fig_fsst_md_10000} shows the case with $n=10^4$. The true value of $\psi$ in our simulations is $\psi_{0}=3$. The identified set is $\Psi_{I}(P) = [2.8, 3.2]$, marked by the vertical dotted lines.
    }\label{fig_fsst_md_comparison}
\end{figure}

In the missing data example, the partially identified parameter vector is the probability vector $P(Y=y,D=d)$ for $y \in \{1,2,3,4,5\}$ and $d \in \{0,1\}$. Thus, our parameter vector has $10$ elements. We then construct confidence sets for the unconditional mean of the variable $Y$ (see Appendix D for additional details). The simulation results are displayed in Table \ref{table_missing_data}. For all sample sizes and confidence levels the simulated coverage probability is above the nominal level. The conservative distortion from our procedure becomes especially obvious at sample sizes larger than $n=10^4$. For samples size $n=10^5$ and $10^6$ the coverage probability is $1$ for all the values of $\alpha$ we consider, although at these sample sizes our confidence set is very close to the identified set. Figure \ref{fig_fsst_md_comparison} shows a comparison of the coverage probabilities of our method versus the methods of \cite{fang2021inference} and \cite{gafarov2021inference} for $n=10^3$ and $n=10^4$. For both sample sizes our proposed procedure is slightly more conservative.

\begin{figure}[!t]
\centering
\begin{subfigure}[!t]{0.49\textwidth}
\centering
\includegraphics[width=\textwidth]{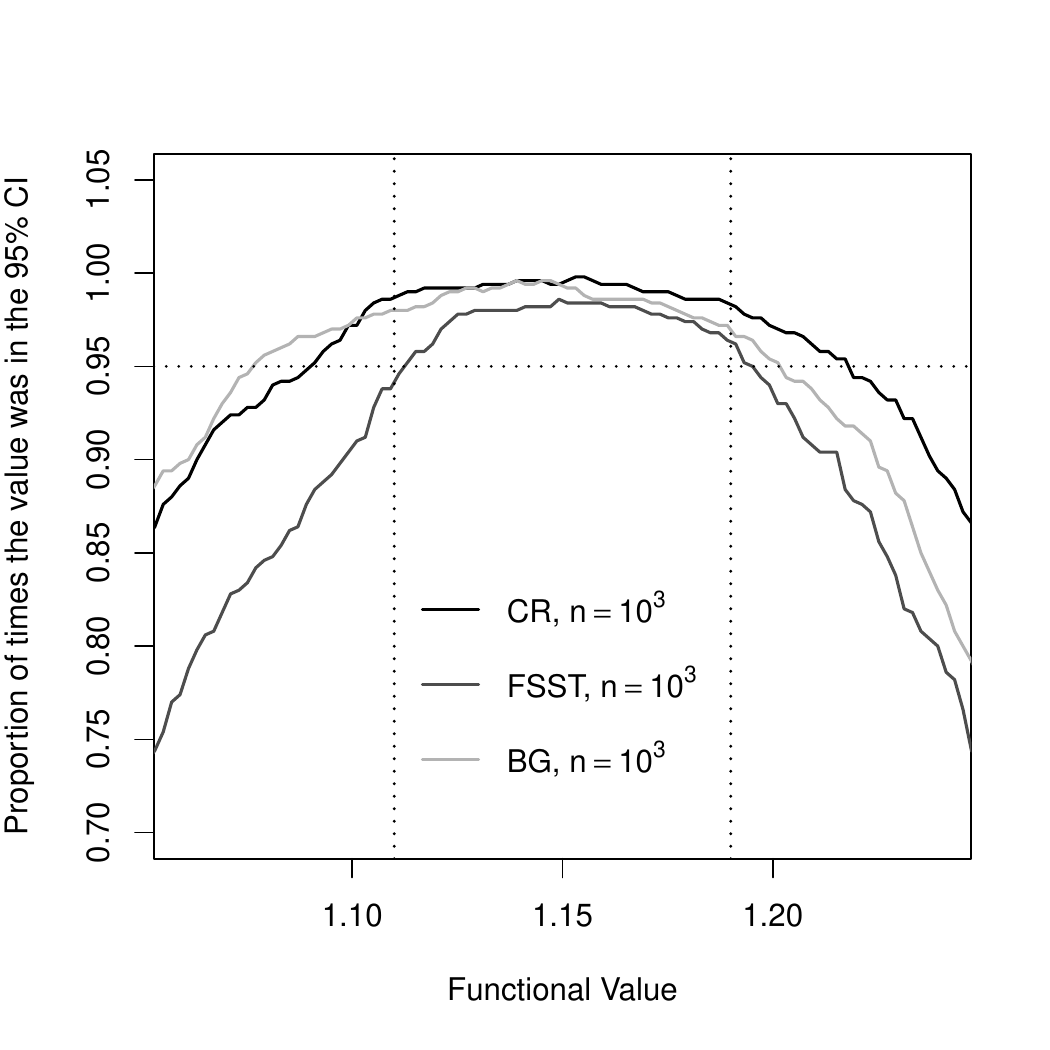}
\caption{} \label{fig_fsst_int_1000}
\end{subfigure}\hfill
\begin{subfigure}[!t]{0.49\textwidth}
\includegraphics[width=\textwidth]{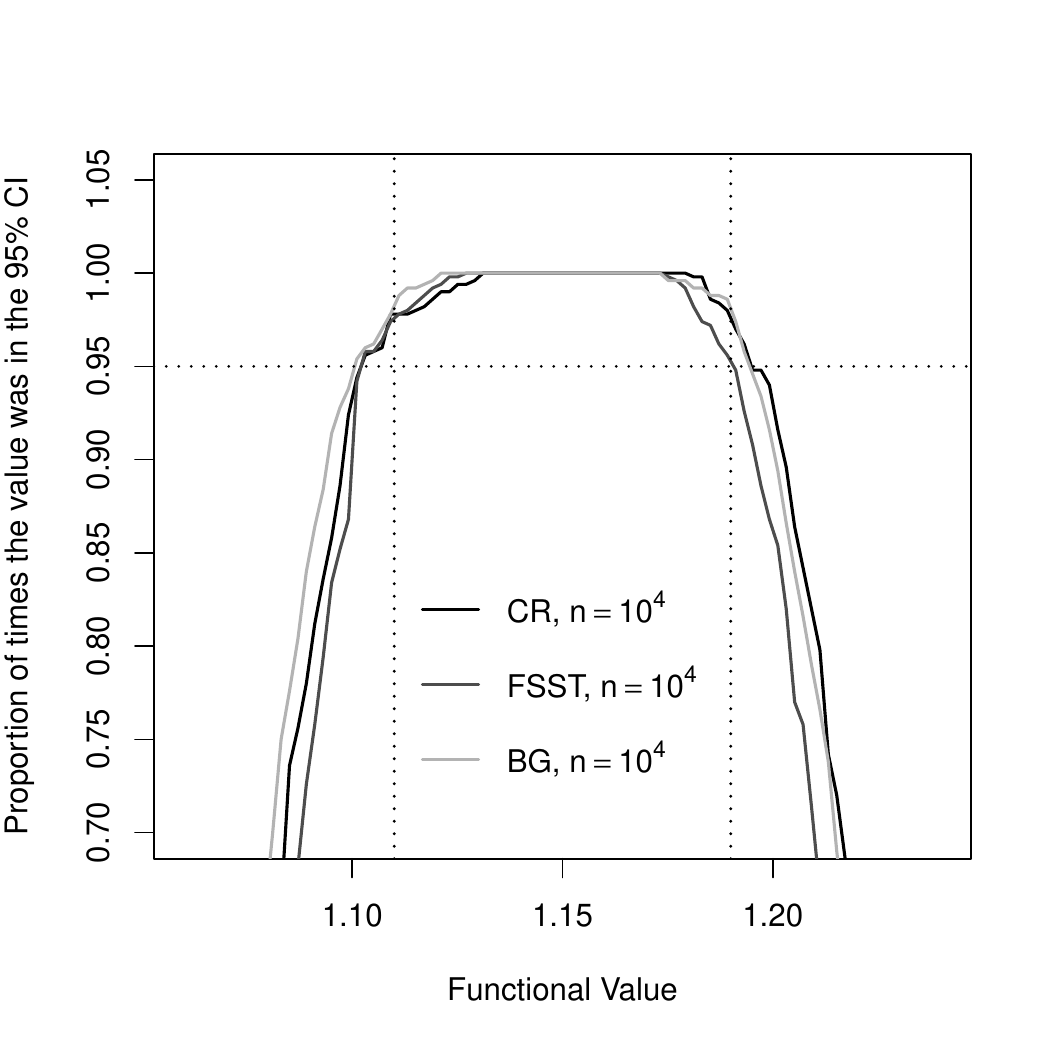}
\caption{} \label{fig_fsst_int_10000}
\end{subfigure}
\caption{
		A comparison between our proposed procedure and the procedure in \cite{fang2021inference} and \cite{gafarov2021inference} for the interval valued regression example. Each subfigure displays the proportion of times various values of the functional $\psi$ associated with the interval valued regression example fell within the $95\%$ confidence set for our proposed procedure (labelled as CR), for the procedure of \cite{fang2021inference} (labelled as FSST), and for the procedure of \cite{gafarov2021inference} (labelled as BG). Figure \ref{fig_fsst_int_1000} shows the case with $n=10^3$, and Figure \ref{fig_fsst_int_10000} shows the case with $n=10^4$. The true value of $\psi$ in our simulations is $\psi_{0}=1.15$. The identified set is $\Psi_{I}(P) = [1.11, 1.19]$, marked by the vertical dotted lines.
    }\label{fig_fsst_int_comparison}
\end{figure}

\begin{figure}[!t]
\centering
\begin{subfigure}[!t]{0.49\textwidth}
\centering
\includegraphics[width=\textwidth]{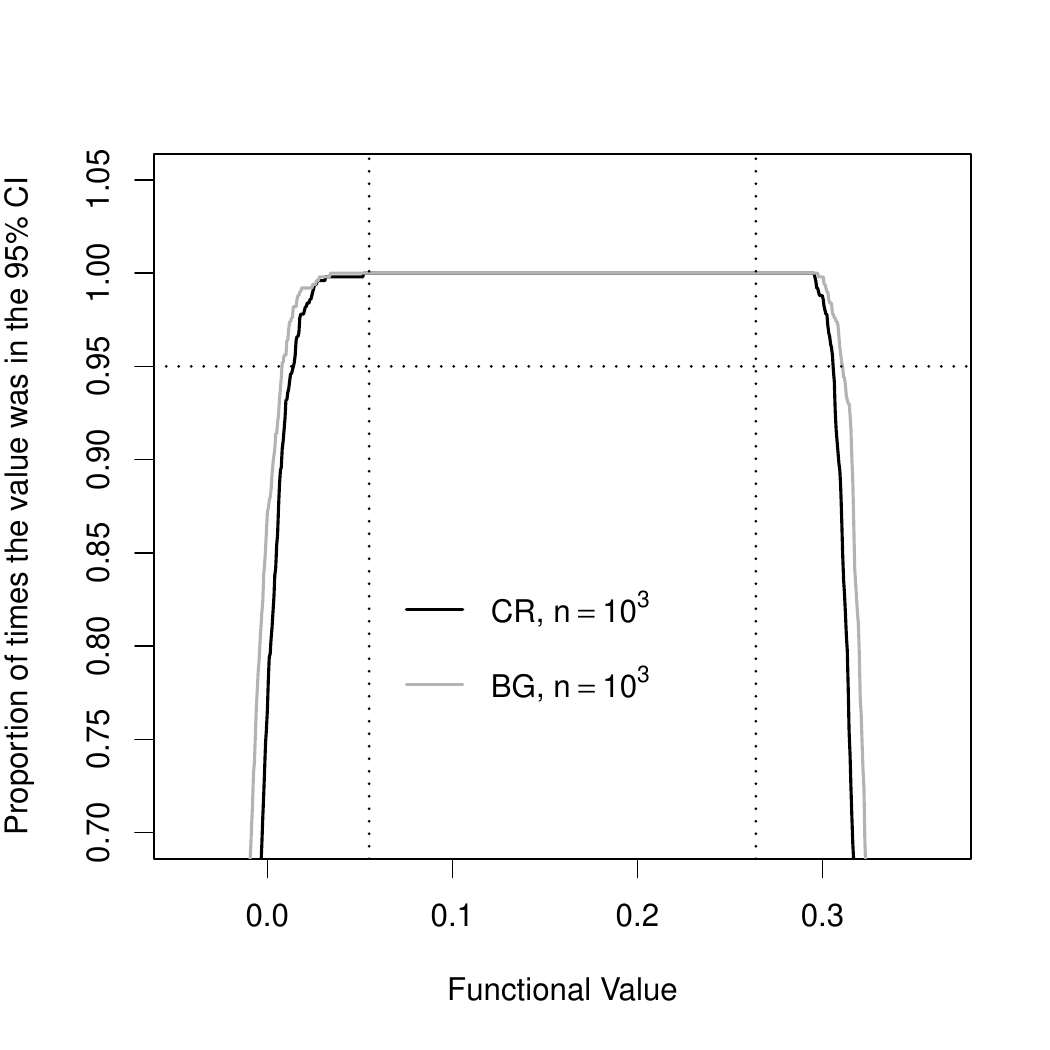}
\caption{} \label{fig_fsst_ct_1000}
\end{subfigure}
\begin{subfigure}[!t]{0.49\textwidth}
\includegraphics[width=\textwidth]{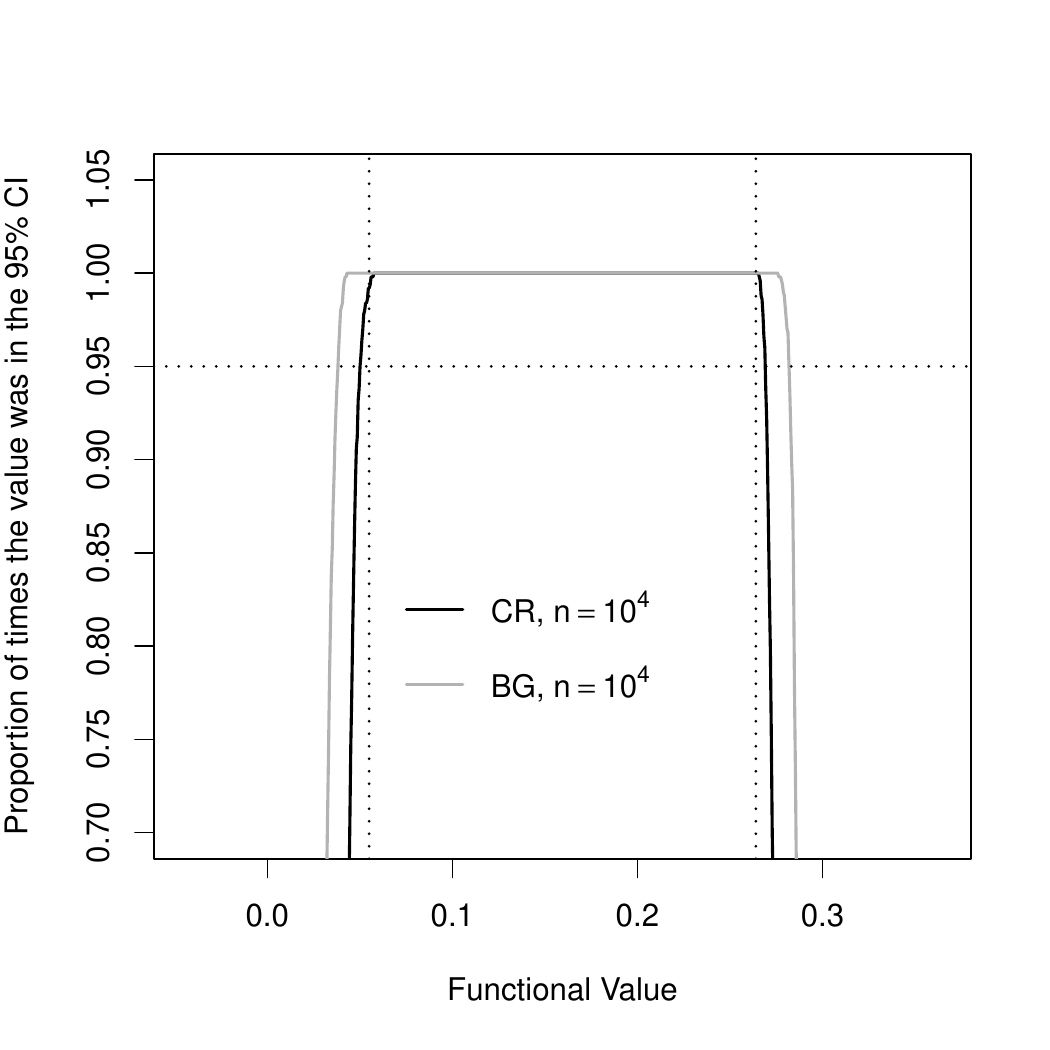}
\caption{} \label{fig_fsst_ct_10000}
\end{subfigure}
\caption{
		A comparison between our proposed procedure and the procedure in \cite{gafarov2021inference} for the counterfactual treatment example. Each subfigure displays the proportion of times various values of the functional $\psi$ associated with the counterfactual treatment example fell within the $95\%$ confidence set for our proposed procedure (labelled as CR), and for the procedure of \cite{gafarov2021inference} (labelled as BG). Figure \ref{fig_fsst_ct_1000} shows the case with $n=10^3$, and Figure \ref{fig_fsst_ct_10000} shows the case with $n=10^4$. The true value of $\psi$ in our simulations is $\psi_{0}=0.216$. The identified set is $\Psi_{I}(P) = [0.054,0.264]$, marked by the vertical dotted lines.
    }\label{fig_fsst_ct_comparison}
\end{figure}

For the interval valued regression example, the partially identified parameter vector is the $4 \times 1$ parameter $\theta$ in the regression $Y=X^\top\theta+\varepsilon$, where $Y$ is interval-valued (see Appendix D for more details). We construct confidence sets for the first component $\theta_{1}$ of this parameter vector. The worst case coverage probability over the identified set is displayed in Table \ref{table_linear_regression_interval}, and is above the nominal level for nearly all sample sizes and confidence levels. Figure \ref{fig_fsst_int_comparison} then shows a comparison of the coverage probabilities of our method versus the method of \cite{fang2021inference} and \cite{gafarov2021inference} at various sample sizes. For $n=10^3$, our proposed procedure has similar performance to the procedure of \cite{gafarov2021inference}, although we find the procedure of \cite{fang2021inference} has more power at values of $\psi$ far from the identified set. At $n=10^4$, the rejection probabilities of the three procedures are quite close.

Finally, the simulation results for the counterfactual treatment example are displayed in Table \ref{table_counterfactual_policy}. This example compares two counterfactual treatment policies. The partially identified parameter is the full joint distribution of $(Y_{0},Y_{1},D)$ conditional on $X$ and $Z$, which has $32$ elements in our DGP. The functional of interest is $\E_{P}[Y^{A}-Y^{B}]$, the difference in expected outcomes under two counterfactual treatment assignments. See Appendix D for details. Similar to the previous examples, the worst case coverage probability over the identified set is above the nominal level for all sample sizes, and the conservative distortion becomes especially apparent at larger sample sizes. Unlike the missing data and interval valued regression examples, this example does not satisfy the assumptions in \cite{fang2021inference} since the functional of interest has a data-dependent gradient. Thus, we only compare with the procedure of \cite{gafarov2021inference}. Our proposed procedure has slightly better power than the procedure of \cite{gafarov2021inference} for this example, although the differences are small. 

\begin{table}[t!]
  \centering
  \caption{Computation time comparison between our proposed procedure and the procedure of \cite{fang2021inference} and \cite{gafarov2021inference}.}
  \begin{footnotesize}
    \begin{tabular}{l|ccc|c}
    \toprule
    \toprule
          & \multicolumn{3}{c|}{Time Per Confidence Set*} &  \\
    \midrule
    Missing Data & $n=250$ & $n=10^3$ & $n=10^4$ & Total Time ($R=500$, $n=10^3$) \\
    \midrule
    BG    & $0.30$ sec & $0.53$ sec & $2.93$ sec & $4.39$ min \\
    CR    & $11.60$ sec & $12.93$ sec & $28.58$ sec & $1.80$ hrs \\
    FSST  & $15.53$ min & $15.82$ min & $20.65$ min & $131.85$ hrs \\
    \midrule
          &       &       &       &  \\
    \midrule
    Interval Valued Regression & $n=250$ & $n=10^3$ & $n=10^4$  & Total Time ($R=500$, $n=10^3$) \\
    \midrule
    BG    & $0.23$ sec & $0.40$ sec & $3.39$ sec & $3.36$ min \\
    CR    & $11.07$ sec & $12.08$ sec & $15.02$ sec & $1.68$ hrs \\
    FSST  & $16.88$ min & $18.66$ min & $39.67$ min & $155.47$ hrs \\
    \midrule
          &       &       &       &  \\
    \midrule
    Counterfactual Treatment & $n=250$ & $n=10^3$ & $n=10^4$  & Total Time ($R=500$, $n=10^3$) \\
    \midrule
    BG    & $1.33$ sec & $2.53$ sec & $16.81$ sec & $21.08$ min \\
    CR    & $13.99$ sec & $16.01$ sec & $40.65$ sec & $2.22$ hrs \\
    FSST  & n/a     &n/a     & n/a     &n/a \\
    \bottomrule
    \bottomrule
    \end{tabular}%
    \end{footnotesize}
        \begin{tablenotes}
      \tiny
      \item $\qquad\qquad$*Average time per confidence set is computed over $500$ replications on one cpu. 
    \end{tablenotes} 
  \label{table_time}%
\end{table}%

Computation times for our proposed procedure and the procedure of \cite{fang2021inference} and \cite{gafarov2021inference} are displayed in Table \ref{table_time}. All times are without any parallelization, which can be used to reduce the computational time of both our proposed method, and the method of \cite{fang2021inference}. The results show that the method of \cite{gafarov2021inference} is the fastest, followed by our method, and then followed by the method of \cite{fang2021inference}.\footnote{Note the method of \cite{gafarov2021inference} does not require any resampling procedure, which explains the computational advantage relative to our method.}

\section{Conclusion}\label{section_conclusion}

This paper proposes a simple procedure for constructing confidence intervals for functionals or scalar subvectors of a partially identified parameter vector. Our confidence interval is constructed by repeatedly solving perturbed bootstrap linear programs, and then selecting the appropriate quantiles of the resulting bootstrap distributions. The efficacy of algorithms for linear programming problems means our confidence set is computationally easy to construct. Uniform validity of our proposed procedure is proven under weak assumptions using genericity results for linear programming problems. We show that our perturbed linear programs are uniformly Hadamard differentiable almost surely with respect to a user-specified perturbation distribution, and use this result to apply a uniform functional delta method. The benefits of the approach are that it is fast, conceptually simple, and valid under weak assumptions. However, the resulting confidence set has a coverage probability tending to 1 over the identified set, and exact coverage on an outer set under some additional assumptions. In this sense, the approach modifies the coverage objective in exchange for some practical benefits. An unresolved question in the paper is how to choose the support of the perturbations in finite samples, although we believe the proposed perturbation approach has enough promise to justify this as a fruitful avenue of future research. ~\\

\noindent \textbf{Disclosure Statement:} The authors report there are no competing interests to declare.

\begin{singlespace}
\bibliographystyle{apa}
\bibliography{bibfile_main}
\end{singlespace}

\end{document}